\documentclass[usenatbib]{mnras}

\usepackage{newtxtext,newtxmath}

\usepackage[T1]{fontenc}

\DeclareRobustCommand{\VAN}[3]{#2}
\let\VANthebibliography\thebibliography
\def\thebibliography{\DeclareRobustCommand{\VAN}[3]{##3}\VANthebibliography}

\usepackage{amsmath,amstext}
\usepackage{graphicx}
\graphicspath{plots}
\usepackage{color}
\usepackage{comment}
\usepackage{hyperref}
\usepackage{calc}
\usepackage{threeparttable}
\usepackage{enumerate}

\usepackage{xcolor}

\def\sfrtrue/{\ensuremath{\mathrm{SFR}_\mathrm{true}}}
\def\sfrind/{\ensuremath{\mathrm{SFR}_\mathrm{ind}}}
\def\sfrfuv/{\ensuremath{\mathrm{SFR}_\mathrm{{FUV}}}}
\def\sfrhalpha/{\ensuremath{\mathrm{SFR}_\mathrm{H\alpha}}}
\def\tmin/{\ensuremath{t\mathrm{_{avg}^{min}}}}
\def\tfifty/{\ensuremath{t_{50\%}}}

\def\halpha/{\ensuremath{\mathrm{{H\alpha}}}}

\usepackage{graphicx}	
\usepackage{amsmath}	

\title[Time-scales of SFR indicators from FIRE SFHs]{The time-scales probed
by star formation rate indicators for realistic, bursty star formation histories from the FIRE simulations}

\date{\vspace{-20pt}Submitted to MNRAS, August 2020}

\pubyear{2020}

\author[Flores Vel\'azquez et al.]{
\parbox{\textwidth}{
Jos\'e A. Flores Vel\'azquez$^{1,2}$,
Alexander B. Gurvich$^{2}$\thanks{Corresponding author. Email: agurvich@u.northwestern.edu},
Claude-Andr\'e Faucher-Gigu\`ere$^{2}$,
James S. Bullock$^{1}$,
Tjitske K. Starkenburg$^{2}$,
Jorge Moreno$^{3}$,
Alexandres Lazar$^{1}$,
Francisco J. Mercado$^{1}$,
Jonathan Stern$^{2}$,
Martin Sparre$^{4,5}$,
Christopher C. Hayward$^{6}$,
Andrew Wetzel$^{7}$,
Kareem El-Badry$^{8}$
}
\vspace{0.4cm}\\
\parbox{\textwidth}{
$^{1}${Department of Physics and Astronomy, University of California, Irvine, CA
92697, USA}\\
$^{2}${Department of Physics \& Astronomy and CIERA, Northwestern University, 1800 Sherman Ave, Evanston, IL 60201, USA}\\
$^{3}$Department of Physics and Astronomy, Pomona College, Claremont, CA 91711, USA\\
$^4$Institut f\"ur Physik und Astronomie, Universit\"at Potsdam, Karl-Liebknecht-Str.\,24/25, 14476 Golm, Germany\\
$^5$Leibniz-Institut f\"ur Astrophysik Potsdam (AIP), An der Sternwarte 16, 14482 Potsdam, Germany\\
$^{6}${Center for Computational Astrophysics, Flatiron Institute, 162 Fifth Avenue, New York, NY 10010, USA}\\
$^{7}${Department of Physics \& Astronomy, University of California, Davis, CA 95616, USA}\\
$^{8}${Department of Astronomy and Theoretical Astrophysics Center, University of California Berkeley, Berkeley, CA 94720, USA}\\
}}
\vspace{-40pt}

\begin{document}
\label{firstpage}
\pagerange{\pageref{firstpage}--\pageref{lastpage}}
\maketitle

\begin{abstract}
    Understanding the rate at which stars form is central to studies of galaxy formation. 
    Observationally, the star formation rates (SFRs) of galaxies are measured using the luminosity in different frequency bands, often under the assumption of a time-steady SFR in the recent past.
    We use star formation histories (SFHs) extracted from cosmological simulations of star-forming galaxies from the FIRE project to analyze the time-scales to which the \halpha/ and far-ultraviolet (FUV) continuum SFR indicators are sensitive. 
    In these simulations, the SFRs are highly time variable for all galaxies at high redshift, and continue to be bursty to $z=0$ in dwarf galaxies. 
    When FIRE SFHs are partitioned into their bursty and time-steady phases, the best-fitting FUV time-scale fluctuates from its ${\sim}10$ Myr value when the SFR is time-steady to ${\gtrsim}100$ Myr immediately following particularly extreme bursts of star formation during the bursty phase. 
    On the other hand, the best-fitting averaging time-scale for \halpha/ is generally insensitive to the SFR variability in the FIRE simulations and remains ${\sim} 5$ Myr at all times. 
    These time-scales are shorter than the $100$ Myr and ${10}$ Myr time-scales sometimes assumed in the literature for FUV and \halpha/, respectively, because while the FUV emission persists for stellar populations older than $100$ Myr, the time-dependent luminosities are strongly dominated by younger stars. 
    Our results confirm that the ratio of SFRs inferred using \halpha/ vs. FUV can be used to probe the burstiness of star formation in galaxies.
\end{abstract}

\vspace{-.4in}

\section{Introduction}
\label{s:introduction}

\begin{figure*}
    \includegraphics[width=\linewidth]{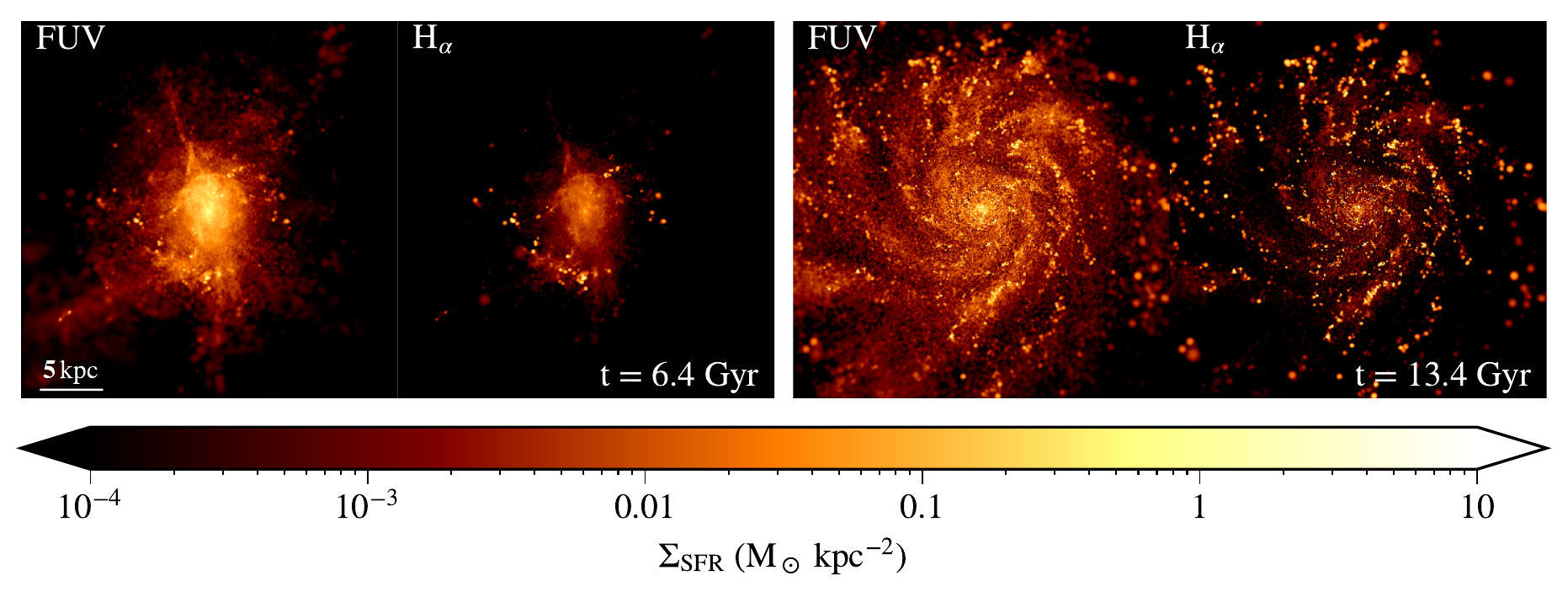}
    \caption{
        \label{f:sfr_maps}
        Star formation rate surface density maps as traced by unattenuated FUV and \halpha/ emission for a simulated Milky Way-mass galaxy (\texttt{m12i}) at cosmic time $t = 6.4$ Gyr ($z=0.9$, ${\approx} 50$ Myr after a burst of star formation; left) and at $t=13.8$ Gyr ($z=0.03$, long after the SFR has settled to a steady state; right).
        The high-redshift panels on the left showcase the irregular morphology of the galaxy during the period of bursty star formation.
        The larger integrated FUV-indicated SFR also shows how the FUV-indicated SFR following a burst of star formation overestimates the more recent SFR indicated by \halpha/.
        In contrast, in the low-redshift panels on the right, which exhibit a well-ordered disk (viewed face-on) after the galaxy's transition to time-steady star formation, have more similar total FUV- and \halpha/-indicated SFRs. 
        In these unattenuated maps, the \halpha/ light is more concentrated in young, bright star-forming regions, while there is more diffuse FUV light between spiral arms.
        }
\end{figure*}

    \subsection{Measuring star formation rates observationally}
    \label{s:measuring_observationally}
        In the regime where counting individual stellar objects and measuring their ages is impossible (e.g. the vast majority of extragalactic systems), star formation rates (SFRs) must be inferred from the observed integrated light from the current stellar population and surrounding ionised gas \citep[e.g.,][]{Kennicutt2012,Calzetti2013}. 
        Two widely used indicators of star formation are the nebular \halpha/ recombination line (6563 \AA) and the far-ultraviolet (FUV) continuum (${\sim}1350-1750$ \AA, for the GALEX bandpass).
        \halpha/ photons, primarily produced in the HII regions surrounding young massive stars, have long been used as an indicator of recent (${\sim} 10$ Myr) star formation due to the short lifetimes of the stars that produce them.
        On the other hand, older (but still relatively young) stellar populations with masses above ${\sim} 5$ M$_\odot$ dominate the integrated spectrum in FUV and are thought to trace star formation on longer-time-scales \citep[up to ${\sim} 100$ Myr;][]{Kennicutt2012,Calzetti2013}.
        SFR indicators can thus be thought of as a convolution of the intrinsic ``true'' star formation history (SFH) with a time-delayed response function.
        One can take advantage of this fact by combining observations of different SFR indicators to measure SFRs averaged over different times in a galaxy's recent history \citep[e.g.,][]{Weisz2012,Dominguez2015,Sparre2017}.
        
        There is a vast literature seeking to predict the properties of simple stellar populations \citep[SSPs, notable examples include][]{Leitherer1999,Bruzual2003,Conroy2009,daSilva2012,Eldridge2017}.
        These SSPs model the evolution of populations of stars from a distribution of zero-age main sequence (ZAMS) masses, known as an initial mass function, or IMF (see \citeauthor{Kroupa1993} \citeyear{Kroupa1993} for one such example). 
        SSPs make predictions for a number of important and useful properties of the populations, including: their spectral energy distributions (SEDs), supernova rates, distributions of stellar type, estimates of metal yields, distributions of surviving stellar masses, distributions of colours,and both ionising photon rates and narrowband FUV continuum emission as a function of age.
        The last two of these are the most relevant for the analysis presented here -- because they are the basis for the SFR response functions mentioned above. 
        
        Once a luminosity response function is determined, the next step is to calibrate a star formation rate indicator by modeling a simple stellar population and the luminosity it emits.
        Thus, calibration requires making some assumption about the shape of the (recent) SFH.
        A common choice is a constant SFR, for which the calibration is the asymptotic luminosity of the population formed at that SFR \citep[e.g.][]{Buat2012,Kennicutt2012}. 
        When suggested by observations (e.g., constrained by ratios of indicators that act on different time-scales) a ``constant burst'' (i.e. a top hat of varying width) is sometimes used \citep[e.g.][]{Calzetti2007,Murphy2011,Hao2011}.
        The appropriate calibration in this case (to recover the burst SFR), is the ratio of the luminosity of the stellar population formed during the burst at the inferred time delay of the observation (typically shortly following the burst) to the burst SFR.
        
        A common definition for the time-scale that an indicator acts on is the time that it takes for the calibration population to reach ${\sim}90\%$ of its integrated luminosity. 
        Following this definition, the FUV emission is thought to trace stars formed over the past ${\sim}100$ Myr \citep{Murphy2011}, whilst \halpha/ emission trace stars formed in the past ${\sim}10$ Myr \citep {Hao2011}.
        However, since even a stellar population formed in a single burst can emit significant UV light long after its FUV or \halpha/ emission peaks, the 90\% ``persistence times'' can be substantially longer than the SFR time-scales to which FUV and \halpha/ light are sensitive. 
        Moreover, the assumption of a constant SFH (or a single top-hat burst) is not necessarily accurate in real galaxies, which have can have complex and often ``bursty'' SFHs. 
        Thus, to properly interpret the SFRs implied by common observational indicators, we must understand the time-scales the indicators probe when applied to realistic SFHs, which may include highly time-variable periods.
        
    \subsection{The need to consider `bursty' star formation}
        \label{s:intro_bursty}
        Deviations in the ``indicated'' (or ``measured'') SFR from the ``true'' instantaneous (or ``intrinsic'') SFR can be substantial when the SFR changes on time-scales shorter than the maximum lifetime of stars that contribute non-negligibly to the observed photons. 
        So-called ``bursty'' star formation is a generic prediction of simulations of galaxy formation that include explicitly resolved star-forming regions and stellar feedback \citep{Governato2012,Teyssier2013,Hopkins2014,Agertz2015,Sparre2017,Faucher-Giguere2018}. 
        Recently, \cite{Iyer2020} characterized the SFR variability using a power spectral density (PSDs) method in several different state-of-the-art galaxy formation models, including hydrodynamic simulations as well as semi-analytic and empirical models. 
        They showed that, while different models predict different PSDs, the hydrodynamic simulations typically predict increasing short time-scale variability with decreasing galaxy mass. 
        The SFR variability arises from a combination of effects, including the interplay between cosmological inflows, mergers, and recycling outflows on longer and intermediate time-scales \citep[e.g.,][]{AA17_cycle}, and processes related to the lifecycle of molecular clouds in the ISM on shorter time scales \citep[e.g.,][]{Tacchella2020}. 
        
        There are also several observational indications that bursty SFRs are important in the real Universe \citep[e.g.][]{Weisz2012,Smit2016,Guo2016,Emami2019,Hirtenstein2019,Pelliccia2020}.
        Specifically, observations use the ratio of short and long time-scale tracers of SFR (e.g. the \halpha/-to-UV ratio) as a signature of bursty star formation \citep[e.g.,][]{Broussard2019}.
        Since immediately after a burst of star formation the short-time-scale tracer decreases faster than the long-time-scale one, a burst leads to a ratio ${<}1$ \citep[unequal dust attenuation could also affect the \halpha/-to-UV ratio and must also be taken into account, e.g.][]{Koyama2019,Salim2020}.
        For example, \citet{Weisz2012} observed a decreasing \halpha/-to-UV ratio with decreasing galaxy mass, suggesting that bursty star formation is more prevalent in dwarf galaxies \citep[these results are echoed by][who use H$\beta$ instead of \halpha/]{Guo2016}. 
        There is also evidence of common SFR variability at high redshift, e.g. in the form of ``extreme emission-line galaxies,'' which have been interpreted as having recently experienced intense starbursts of duration ${\sim} 15$ Myr \citep[][]{vdW2011, Forrest2017}.
        
\begin{figure*}
    \includegraphics[width=\linewidth]{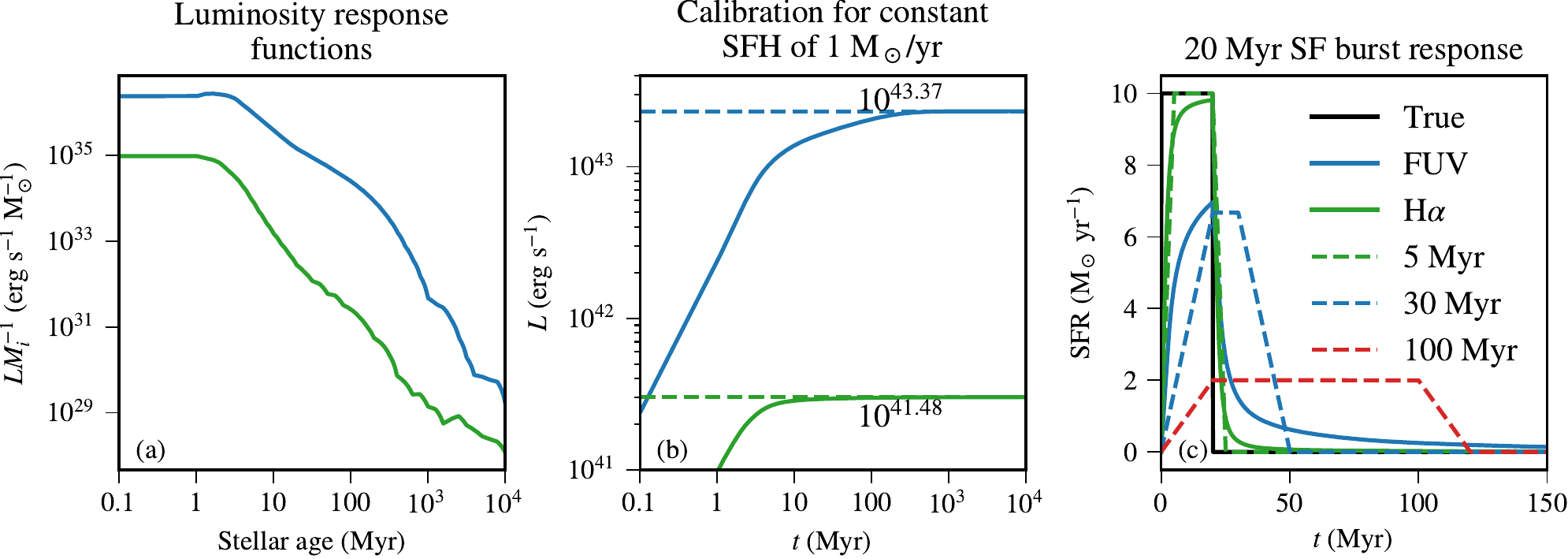}
    \caption{
        \label{f:indicators}
        \textit{(a):} Luminosity (per stellar mass formed) following a star formation burst in the FUV continuum and \halpha/ as a function of stellar age, based on a \texttt{BPASS} stellar population synthesis model. 
        \textit{(b):} Evolution of $L_{\mathrm{FUV}}$ and $L_{\mathrm{H}\alpha}$ as a function of time assuming a SFH with a constant SFR of 1 $\mathrm{M_\odot~ yr^{-1}}$.
        SFR indicator calibration constants are defined as the asymptotic luminosity in each case, represented by dashed lines and annotated on each curve. 
        \textit{(c):} Indicated SFR response to a population of stars formed in a 20 Myr burst (solid black line). 
        Coloured dashed lines show moving boxcar averages of the true SFR using different widths: 5 Myr (green), 30 Myr (blue), and 100 Myr (red). }
\end{figure*}
        Understanding the prevalence and properties of bursty SFHs is of major importance for understand how galaxies form and evolve. 
        The $\mathrm{SFR}-{\rm stellar~mass}$ ($M_{*}$) relation \citep[also known as the ``star formation main sequence'';][]{Brinchmann2004,Noeske2007,Elbaz2007,Whitaker2012,Salim2014,Shivaei2015,Schreiber2015} suggests that galaxies form most of their stars in a steady process. 
        However, as summarized above, simulations and observations both indicate that many galaxies undergo short starbursts, during which they lie well above this relation \citep[see e.g.][]{Willett2015,Sparre2017,Elbaz2018, Matthee2019}. 
        In detail, it is therefore possible that many galaxies frequently move above and below the mean $\mathrm{SFR}-M_{*}$ relation, rather than evolving smoothly along it \citep[e.g.][]{Orr2017}. 
        The scatter in the observed relation should then reflect, in part, the burstiness of star formation in galaxies \citep[e.g.,][]{Caplar2019}. 
        In particular, \cite{Sparre2017} compares the scatter in the observed $\mathrm{SFR}-M_{*}$ relation to simulations from the FIRE (Feedback in Realistic Environments) project\footnote{FIRE project website: \url{http://fire.northwestern.edu}} and find that, to first order, the time-variability of the SFHs in the simulations is consistent with the observed scatter in $\mathrm{SFR}-M_{*}$ (although some quantitative differences may remain to be explained).

        Understanding the time-scales probed by SFR indicators is also relevant for observations outside the UV and for the interpretation of galaxies expected to deviate strongly from the main sequence, e.g. during or following major mergers.
        Using simulations of merging galaxies combined with detailed radiative transfer, \cite{Hayward2014} showed that the total infrared (IR) luminosity (when interpreted using standard assumptions)\footnote{Since IR light is produced by reprocessing of UV light by dust, it is expected to probe time-scales similar to FUV.} 
        over-estimates the true SFR immediately after the merger-induced star formation burst. 
         For major mergers in which a strong starburst is induced, the SFR inferred from the IR luminosity can overestimate the instantaneous SFR during the post-starburst phase by greater than two orders of magnitude.

    \subsection{This paper}
    \label{sec:this_paper}
        In this paper, we ask the following questions: (1) What are the SFR time-scales probed by \halpha/ and FUV indicators when taking into account realistic SFHs; and (2) how do the results depend on SFR burstiness? 
        To answer these questions, we use SFHs from the FIRE-2 simulations of galaxy formation. 
        We also use the detailed datasets provided by the simulations to assess how the indicated SFR time-scales depend on redshift and galaxy mass.
        Since our analysis models the application of standard SFR indicators on the simulated SFHs, the results will inform the interpretation of both past and future observational SFR measurements.
        
        The FIRE-2 simulations are the second generation cosmological, zoom-in hydrodynamic simulations from the FIRE project \citep[][]{Hopkins2014, Hopkins2018}. 
        These high-resolution simulations include a detailed model for star formation and stellar feedback and produce galaxies that match a number of observed properties, which motivates their use as models for ``realistic'' SFHs. 
        The properties include the $M_*-{\rm halo mass}$ ($M_h$ relation), mass-metallicity relations \citep[][]{Ma2016_MZR}, the Kennicutt-Schmidt (KS) relation \citep{Hopkins2014,Orr2018}, and constraints on galactic winds from observations of the circumgalactic medium \citep[e.g.,][]{FG15,FG16,Hafen2017}.
        
        Notably for our purposes, the FIRE simulations predict two main phases of star formation: bursty and time steady \citep[e.g.,][]{Muratov2015,Faucher-Giguere2018}.
        At low redshift, ${\sim}$ Milky Way-mass galaxies form long-lived stable disks with time-steady SFR \citep[e.g.,][]{Gurvich2020}. 
        At high redshift, the progenitors of these galaxies have much more irregular morphologies and time-variable star formation histories \citep[see e.g.][]{Sparre2017,Ma2017_highz_gradients,Stern2020}.
        Figure \ref{f:sfr_maps} illustrates the different morphologies predicted in the bursty and steady phases through the lens of the SFR indicators (FUV and \halpha/ light, uncorrected for dust obscuration).
        The left panels ($z=0.9$) feature a clumpy and amorphous distribution of star formation while the right panels ($z\sim0$) show coherent spiral patterns in the same galaxy but at a later time. 
        In the FIRE simulations, star formation typically remains bursty until $z=0$ in dwarf galaxies.
        
        Although the present study is an important step in using detailed simulations to better understand the time-scales probed by different SFR indicators, we note a few limitations that should be borne in mind when using our results to interpret observations. 
        First, as demonstrated quantitatively in the PSD comparison study by \cite{Iyer2020}, different simulations predict different SFR variability statistics. 
        Thus, in the bursty regimes, our results will apply specifically to the galaxy formation physics in the FIRE-2 simulations (including the stellar feedback models described in more detail below). 
        Second, other indicators of SFR variability than the H$\alpha$-to-FUV ratio that we model in this work can and have been used to empirically constrain SFR burstiness, including the 4000~\AA~break and H$\delta_{\rm A}$ indices \citep[e.g.,][]{Kauffmann2014}. 
        It would be interesting to use simulations to also model these other indicators in the future. 
        Third, we must stress an important caveat regarding effects of dust, discussed in more detail in \S \ref{sec:dust_attenuation}. 
        Dust will in general attenuate FUV and H$\alpha$ light differently, as well as in ways that correlate with SFR variations. 
        Thus, variations in observed FUV and H$\alpha$ light curves can follow substantially different patterns than the intrinsic star formation history. 
        In this work, we do not model dust attenuation but instead assume that observational techniques are able accurately correct for dust and infer the non-attenuated FUV and H$\alpha$ fluxes.

        The remainder of the paper is organised as follows. 
        In \S \ref{s:methods} we describe the simulations in more detail, as well as how we define and analyze SFR indicators. 
        In \S \ref{s:results} we infer the best-fitting averaging time-scales as a function of cosmic time in galaxies of different masses. 
        We discuss our results in \S \ref{s:discussion} and in \S \ref{s:conclusion} we summarise and conclude.

\section{Methods}
\label{s:methods}
\subsection{Star formation histories from FIRE-2 simulations}
            \label{s:SFHs}
            The FIRE-2 simulations are cosmological zoom-in simulations run in the Meshless Finite Mass (MFM) mode of the GIZMO gravity+magnetohydrodynamic code.\footnote{Information about GIZMO and a public version of the code are available at: \url{https://www.tapir.caltech.edu/~phopkins/Site/GIZMO}.} 
            MFM is a Lagrangian, mesh-free, finite-mass method combining the advantages of traditional smooth particle hydrodynamics and grid-based methods
            \citep[for numerical details and tests, see][]{Hopkins2015}. 
            The FIRE-2 physics model, which is described in full detail in \cite{Hopkins2018}, 
            includes radiative cooling for gas down to $10$ K (including approximate treatments of fine-structure metal and molecular lines).  
            Star particles representing simple stellar populations are formed in gas that is self-gravitating, dense ($n_\mathrm{H}\geq 1000$ cm$^{-3}$), and molecular \citep[][]{Hopkins2013}. 
            Star particles return mass, metals, momentum, and energy in the interstellar medium (ISM), at rates that are functions of the star particle's age following the \texttt{STARBURST99} population synthesis model \citep{Leitherer1999}. 
            Stellar feedback is modeled on the scale of star-forming regions and includes  supernovae (Type II and Ia), stellar winds from O, B, and AGB stars, photoelectric heating, photoionization, and radiation pressure. 

            We study analyze a range of galaxy masses, from dwarf galaxies up to ${\sim} L^{\star}$. 
            This allows us to study the different types of SFHs experienced by star-forming galaxies. 
            We analyze eight simulations in total: four dwarfs with $z=0$ halo mass in the range $M_h \sim 10^{9} - 10^{11}$ M$_{\odot}$ 
            and four galaxies with halo mass $M_h \sim 10^{12}$, similar to the Milky Way. 
            Table \ref{t:sim_table} catalogs some basic physical properties of these galaxies at redshift $z=0$. 
            The mass of baryonic resolution elements ranges from $m_{\rm b}\approx 250$ $M_{\odot}$ (for dwarfs) to ${\approx}7100$ $M_{\odot}$ (for the m12 galaxies). 
            More details on resolution and other facets of the FIRE-2 simulations are given in \cite{Hopkins2018} and \cite{Garrison-Kimmel2017,Garrison-Kimmel2019}. 
            All simulations include a sub-grid model for turbulent metal diffusion between nearby gas cells.

            We use an ``archaeological'' approach to reconstruct the SFHs of the galaxies in the FIRE simulations at high time resolution.
            Specifically, we first identify all of the stars in the galaxy's main halo at redshift $z=0$. 
            To do so, we use the AMIGA halo finder \citep{Gill2004,Knollmann2009} to locate the main halo in our simulation volume and measure the virial radius ($R_\mathrm{vir}$) and the halo centre of mass (measured using all of the dark, gas, and stellar components of the galaxy).
            Dark matter haloes are defined using the \citet{Bryan1998} redshift-dependent overdensity criterion.
            We extract the star particles within a spherical volume of radius $5 R_{*,\mathrm{half}}$, where $R_{*,\mathrm{half}}$ is the radius containing 50\% of the stars inside $0.15 R_\mathrm{vir}$.
            The time-dependent SFH is then reconstructed from the distribution of star particle formation times (SFTs) and star masses, self-consistently accounting for stellar mass loss. 
            This archaeological approach neglects the fact that some stars formed in galaxies other than the main progenitor (and then merged), but this is a good approximation for galaxies up to ${\sim} L^{\star}$, which form most of their stars \emph{in situ} \citep[e.g.,][]{AA17_cycle, Fitts18}. 
            
\begin{figure}
    \centering
    \includegraphics[width=\linewidth]{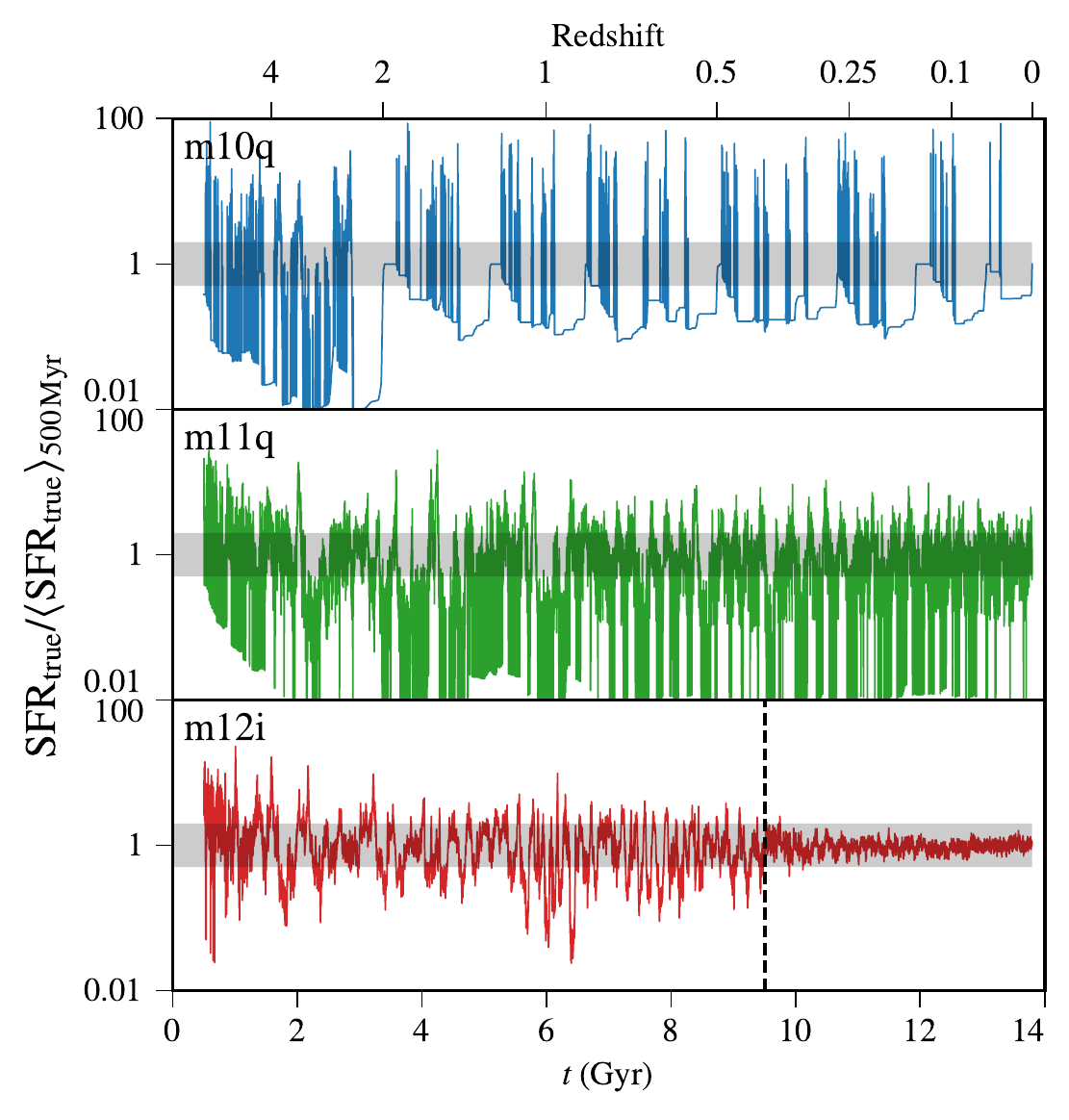}
    \caption{\label{f:cosmic_SFH} 
    Star formation histories of three prototypical galaxies from our simulation sample, one from each halo mass bin considered (\texttt{m10q}, halo mass $M_h \sim 10^{10}$ M$_\odot$ at $z=0$; \texttt{m11q}, $M_h \sim 10^{11}$ M$_\odot$; \texttt{m12f}, $M_h \sim 10^{12}$ M$_\odot$). 
    In each panel, the SFR is normalized by the 500 Myr running boxcar-averaged SFR.
    Dwarf galaxies (top two panels) have bursty SFHs at all redshifts. 
    Milky Way-mass galaxies with $M_h(z=0) \sim 10^{12}$ M$_\odot$ (bottom panel) undergo a transition from a bursty SFR at high redshift to a time-steady SFR at low redshift (with some remaining, but much smaller fluctuations). 
    The transition from bursty to steady SFR is indicated by the vertical black dashed line in the bottom panel.
    Shaded regions show a factor of 0.3 dex in either direction of unity.
    }
\end{figure}

\begin{table}
    \caption{\label{t:sim_table} Properties at $z=0$ of the simulated galaxies analyzed in this paper.}
    \centering
    \begin{threeparttable}
        \centering
        \begin{tabular}{c|c|c|c|c}
          Name & 
          \,\,\,\,$M_h$ ($M_\odot$)\tnote{a}\,\,\,\, & 
          \,\,\,$M_*$ ($M_\odot$)\tnote{b}\,\,\, & 
          \,\,\,$R_{*,\mathrm{half}}$ (kpc)\tnote{c}\,\,\,  &
          \,\,\,$f_g$\tnote{d}\,\,\,  \\  
          \hline
          \hline
m10q & $7.7\times 10^{9}$ & $2.3\times 10^{6}$ & 0.8 & 0.68\\
m11i & $6.8\times 10^{10}$ & $1.0\times 10^{9}$ & 3.7 & 0.61\\
m11q & $1.4\times 10^{11}$ & $6.5\times 10^{8}$ & 2.5 & 0.67\\
m11d & $2.7\times 10^{11}$ & $4.5\times 10^{9}$ & 6.9 & 0.57\\
m12i & $9.4\times 10^{11}$ & $6.8\times 10^{10}$ & 2.9 & 0.20\\
m12b & $1.1\times 10^{12}$ & $9.0\times 10^{10}$ & 2.8 & 0.16\\
m12c & $1.1\times 10^{12}$ & $6.4\times 10^{10}$ & 3.4 & 0.22\\
m12f & $1.3\times 10^{12}$ & $8.8\times 10^{10}$ & 4.0 & 0.24\\
\hline
        \end{tabular}
        \begin{tablenotes}
            \item[a] Total dark matter and baryonic mass within $R_\mathrm{vir}$, the virial radius defined as in \cite{Bryan1998}. 
            \item[b] Total stellar mass within 5$R_{*,\mathrm{half}}$.
            \item[c] 3D radius containing half of the total stellar mass within 15\% of $R_\mathrm{vir}$.
            \item[d] Gas fraction $f_\mathrm{g} = M_\mathrm{g}/(M_*+M_\mathrm{g})$, where $M_\mathrm{g}$ is the gas mass within the galaxy.
        \end{tablenotes}
        \end{threeparttable}
    \end{table}
            From this high time-resolution SFH, we can construct SFHs with the SFR boxcar-averaged on an arbitrary time-scale $t_\mathrm{avg}$, which we denote $\langle \mathrm{SFR_{true}}\rangle_{t_\mathrm{avg}}(t)$. 
            Since, in observational applications, only the SFR before the time of observation is well defined, we define these boxcar averages by averaging the \emph{preceding} $t_\mathrm{avg}$ time interval preceding a given cosmic time $t$.  
            Figure \ref{f:cosmic_SFH} shows the SFH for one representative simulation in each mass bin. 
            
            For each galaxy, we identify the ``end'' of bursty star formation by computing the root mean square relative error (RMSRE) in log-space using a 500 Myr-wide moving boxcar :
            \begin{equation}
                \mathrm{RMSRE_{SFH}} = \sqrt{\frac{\langle\log_{10}(\mathrm{SFR})^2 \rangle_\mathrm{500~Myr} - \langle \log_{10}(\mathrm{SFR})\rangle_\mathrm{500~Myr}^2}{\langle \log_{10}(\mathrm{SFR})\rangle_\mathrm{500~Myr}^2}}.
            \end{equation}
            To account for periods when the SFR is identically zero, we add one tenth of the minimum true SFR to the entire SFH (this choice does not affect any of our results).
            We consider the SFH to become ``steady'' when RMSRE$_\mathrm{SFH} \leq 0.3$. 
            The bottom panel of Figure \ref{f:cosmic_SFH} shows an example of this metric applied to the \texttt{m12i} simulation and how it accurately identifies the clear change in SFR behavior in this case. 
            Dwarf galaxies in FIRE are typically bursty all the way to $z=0$, so no transition is identified for these galaxies.
            
    \subsection{Star formation rate indicators}
        \label{s:indicators}
        We model stellar spectra as a function of age using version 2.2.1 of the Binary Population and Spectral Synthesis (\texttt{BPASS}) stellar population synthesis code \citep{Eldridge2017}. 
        \texttt{BPASS} models the effects of binary evolution on stellar spectra and uses CLOUDY \citep{Ferland1998} to compute nebular emissions, which is the main contributor to \halpha/ emission.\footnote{The \texttt{STARBURST99} stellar population synthesis model used for the FIRE feedback model does not include binaries. As in previous work with BPASS \cite[e.g.,][]{Ma16_binaries}, we neglect possible dynamical differences (during the hydrodynamic simulations) between the different stellar evolution models but include the spectral differences which are most directly relevant to the present study.}
        We use a fixed \texttt{BPASS} model, which consists of a Kroupa IMF \citep{Kroupa1993} with a metallic mass fraction of $Z = 0.014$. 
        To derive observationally-indicated SFR values, $\mathrm{SFR}_{\mathrm{FUV}}$ and $\mathrm{SFR}_{\mathrm{H}\alpha}$, we average the specific BPASS FUV flux over the wavelength interval 1556-1576 \AA~(near the center of the GALEX FUV channel) and use the \halpha/ flux modeled by BPASS. 
        This results in stellar age-dependent ``response functions,'' shown in Figure \ref{f:indicators}a. 

        In order to convert the modeled (``observed'') luminosity into an indicated star formation rate, we assume a constant light-to-SFR ratio, i.e.
        
        \begin{equation} 
            \label{e:light_to_SFR}
            \mathrm{SFR}_\mathrm{ind} = C^{-1}L_\mathrm{ind}
        \end{equation} 

        \noindent where $C$ is the SFR ``calibration constant.''
        This constant can be determined by considering the asymptotic luminosity, $L_\mathrm{{ind}}(t\to\infty)$, of a population of stars forming at a constant rate of $\mathrm{SFR_{ind}}$. 
        These calibration constants neglect obscuration by dust, which should be accounted for when attempting to measure the SFR of real galaxies. 
        In principle, one can relax the assumption of time-constant SFR. 
        In \S \ref{s:indicated_errors} we quantify the expected errors in indicated SFRs (relative to true SFRs), assuming that real SFHs are well modeled by the FIRE-2 simulations we analyze. 
        
        Figure \ref{f:indicators}b shows $L_\mathrm{{ind}}$ as a function of time for $\mathrm{SFR_{true}}(t) = 1 ~\mathrm{M_\odot~ yr^{-1}}$. 
        For the FUV continuum, $L_\mathrm{ind}=\nu \langle L_{\nu} \rangle$ is evaluated at the middle of the FUV wavelength sample (1566~\AA). 
        Using Equation (\ref{e:light_to_SFR}), we find calibration constants of $C_\mathrm{{FUV}} = 10^{43.33}~ \mathrm{~erg ~s^{-1}/M_\odot~yr^{-1}}$ and $C_\mathrm{{H\alpha}} = 10^{41.42} ~ \mathrm{~erg ~s^{-1}/M_\odot~yr^{-1}}$.    
        We note that each indicator reaches $L_\mathrm{ind} \simeq 0.9 L_\mathrm{ind}(t\to \infty)$ at $t \approx 100$ Myr and $t \approx 10$ Myr respectively-- the often quoted time-scales for each of these indicators.
        
        Figure \ref{f:indicators}c shows the SFR response of these indicators applied to a population of stars formed by a 20 Myr burst of star formation at 10 M$_\odot$ yr$^{-1}$, beginning at $t=0$. 
        We compare the indicated SFRs to three moving ``boxcar'' averages of width 5, 30, and 100 Myr.
        One important takeaway is that the 100 Myr running average is a poor match to the FUV indicated SFR at all times, which hints that the ``effective'' FUV response time-scale is much shorter. 
        On the other hand, the 5 Myr average agrees well with the \halpha/ indicated SFR. 

\begin{figure*}
    \includegraphics[width=\linewidth]{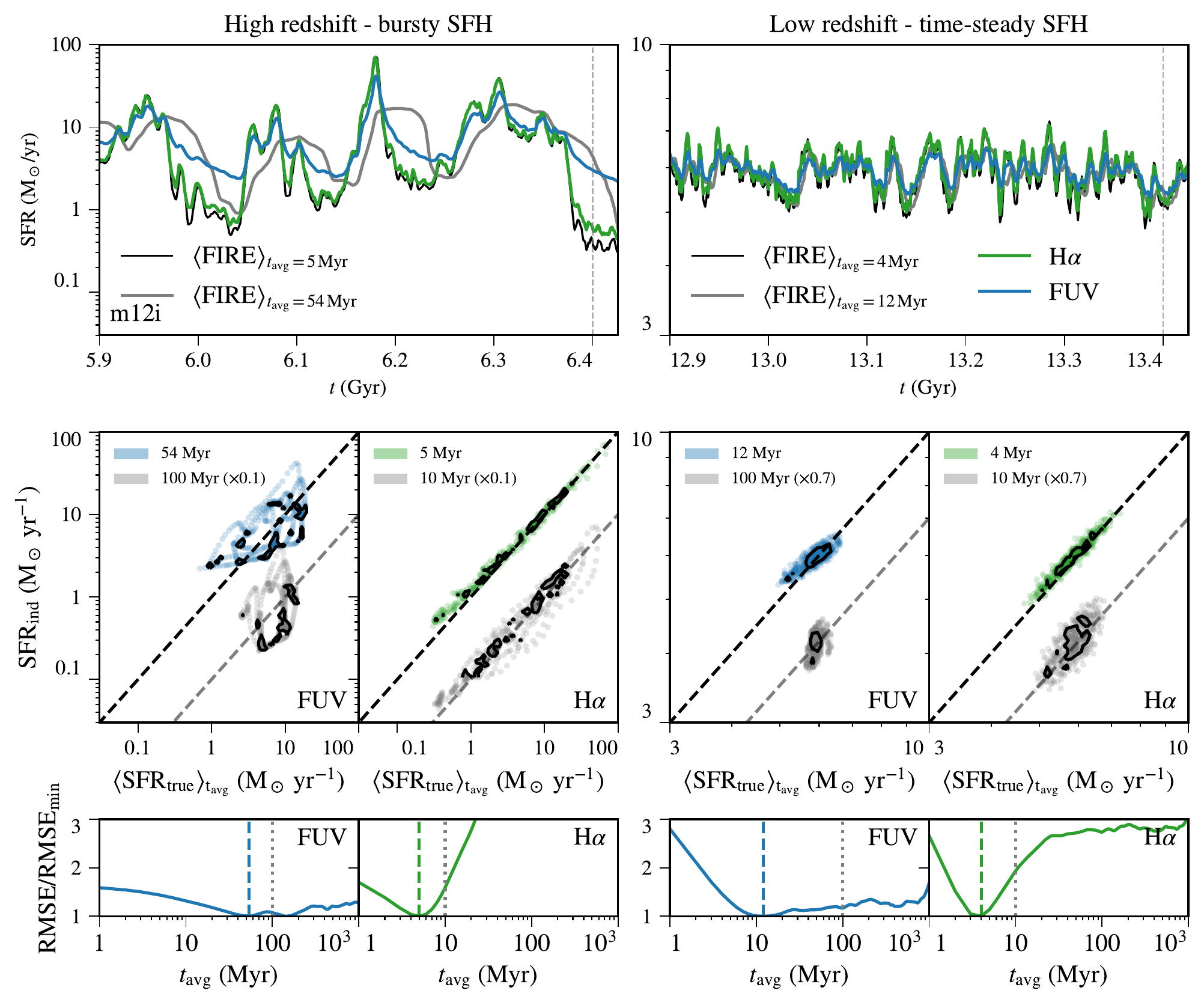}
    \caption{
        \label{f:scatter_summary}
        \textit{Top:} ``True'' and ``indicated'' SFHs in a 500-Myr window at high (left) and low (right) redshift in \texttt{m12i}.
        True SFHs are plotted using best-fitting, boxcar-average equivalent time-scales for FUV (grey) and \halpha/ (black).
        Grey vertical lines are plotted to identify the times corresponding to the renderings in Figure \ref{f:sfr_maps}.
        At early times the SFR frequently changes by a factor of $>10$ on time-scales ${\sim}30-50$ Myr while at late times it varies only by ${\lesssim}10\%$.
        \textit{Middle:} Scatter plot of the indicated star formation rate, SFR$_\mathrm{ind}$, vs. the true star formation rate, boxcar-averaged using the best-fitting averaging time-scale (coloured). 
        Contours containing 50\% of the points are plotted in black alongside a dashed line showing along the one-to-one relation. 
        For comparison, scatter plots for SFR boxcar-averaged over time-scales of 100 Myr and 10 Myr are also shown in grey in the FUV and \halpha/ panels, respectively (offset vertically by the factors noted on the figure for visual clarity). 
        The smaller scatter in the colored points vs. the grey is visible by eye in all four panels. 
        \textit{Bottom:} Root mean square error (RMSE) between the indicated and time-averaged true SFHs as a function of boxcar averaging time-scale $t_\mathrm{avg}$, normalized by the minimum RMSE.
        Vertical dashed lines (coloured) are plotted at the locations of the minima, while the dotted (grey) vertical lines indicate 100 Myr and 10 Myr time-scales for reference.
        At both early and late times, \halpha/ has a well-defined minimum RMSE in almost all time windows. 
        For FUV, on the other hand, the RMSE curve is typically much shallower, especially in the bursty high-redshift regime, implying that the equivalent boxcar average time-scale is not as well defined. 
        Note that the value of \tmin/ inferred in the high-redshift window shown here is not necessarily representative since the best-fitting time-scale fluctuates strongly following particularly extreme bursts of star formation like the one shown here (see Figure \ref{f:tmin_history}).
        }
\end{figure*}
        \subsection{Inferring the time-scales of SFR indicators}
            \label{s:scatter}
            Next, we compare the true SFH to the ``indicated'' SFHs using the calibrated indicators from \S \ref{s:indicators}.
            To measure the indicated SFR$_\mathrm{ind}(t)$ (i.e., the SFR that would be inferred using a standard observational indicator, as defined in the previous section), we consider the total luminosity from the stars in a galaxy with SFT$<t$, i.e.
            \begin{equation}
                \label{e:sfr_ind_methods}
                \mathrm{SFR_{ind}}(t) = C_\mathrm{ind}^{-1} \sum_{i,~\mathrm{SFT}_i < t} L^i_\mathrm{ind}(t-\mathrm{SFT}_i),
            \end{equation}
            \noindent where $L^i_\mathrm{ind}(t-\mathrm{SFT}_i)$ corresponds to the response functions in Figure \ref{f:indicators}a, evaluated for star particle $i$ with a stellar age $(t-\mathrm{SFT}_i)$.
            
            Our methodology for comparing the true and indicated SFRs is illustrated in Figure \ref{f:scatter_summary}. 
            At any given time there is, in general, a non-zero deviation between the true and indicated SFRs
            The goal of our analysis is to identify the \textit{best-fitting boxcar averaging time-scale}, \tmin/, that minimises this difference between $\langle \sfrtrue/\rangle_{t_\mathrm{avg}}$ and a given SFR$_\mathrm{ind}$. 
            Physically, \tmin/ is the width of the boxcar over which one would have to average the true SFR to match the indicated SFR at the time of an observation.
            \begin{figure*}
    \includegraphics[width=\linewidth]{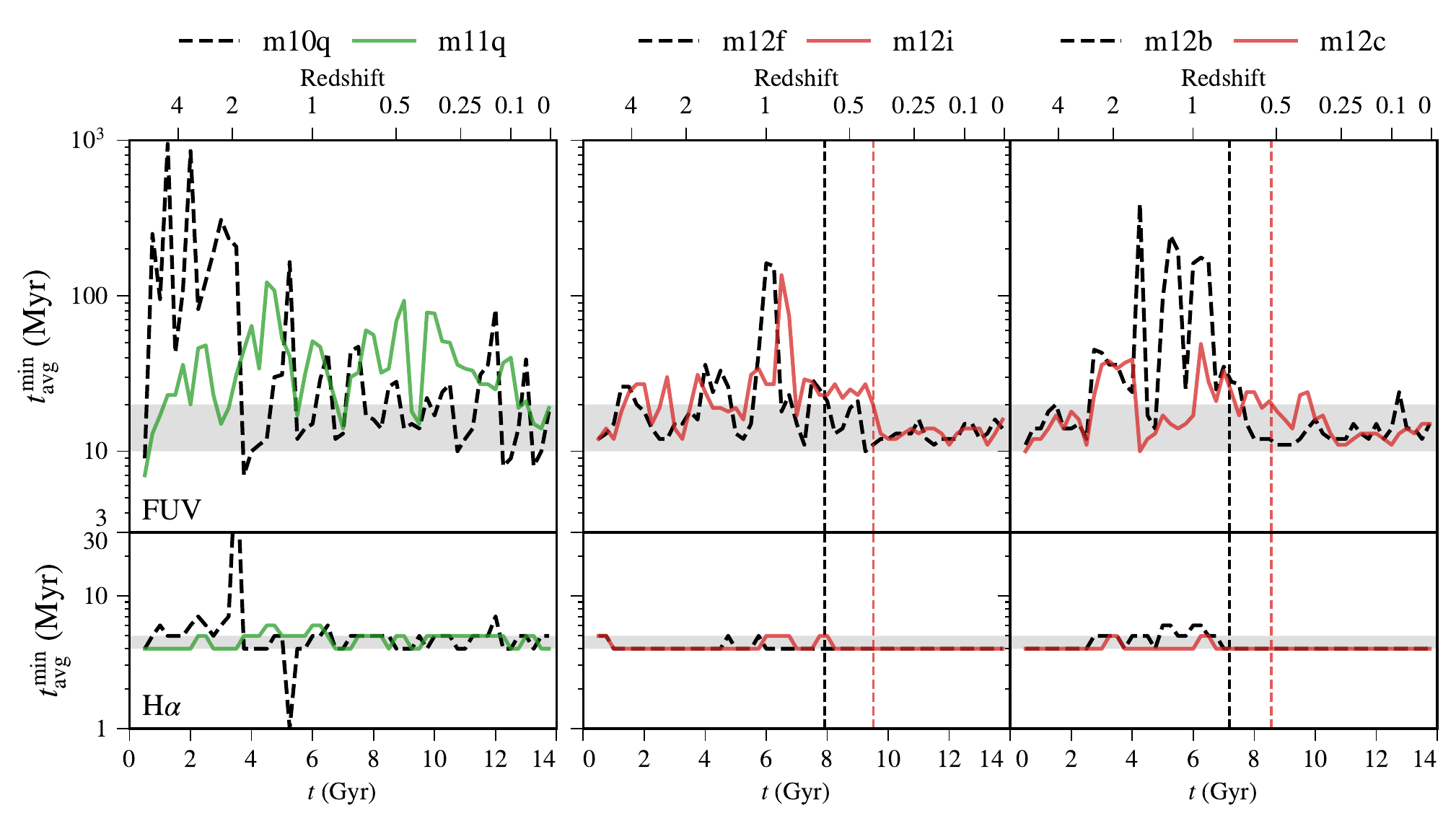}
    \caption{
        \label{f:tmin_history}
        Best-fitting boxcar-average equivalent time-scale, \tmin/, for both FUV (top) and \halpha/ (bottom) for different simulations as a function of cosmic time.
        Grey shaded regions are shown to help identify departures from the time-scales we identify in the time-steady limit for each indicator ($10-20$ Myr for FUV and $4-5$ Myr for \halpha/).
        \textit{Left}: Simulations with halo mass $M_h \sim 10^{10} ~\mathrm{M_\odot}$ and $M_h \sim 10^{11}~ \mathrm{M_\odot}$ at $z=0$. 
        \textit{Middle and Right}: Simulations with $M_h \sim 10^{12} ~\mathrm{M_\odot}$.
        The simulations of Milky Way-mass galaxies with $M_h \sim 10^{12} ~\mathrm{M_\odot}$ experience a transition from bursty to time-steady SFR at the time indicated by the vertical dashed lines.
        For time periods when the SFR is bursty (regardless of halo mass) the best-fitting averaging time-scale for FUV fluctuates strongly and reaches much higher values than for time periods when the SFR is time-steady. 
        For \halpha/, the shorter best-fitting averaging time-scale $4-5$ Myr is generally stable and insensitive to the shape of the SFH. 
         }
\end{figure*}
            
            When the SFH fluctuates rapidly, many $t_\mathrm{avg}$ may produce the same $\langle \sfrtrue/\rangle_{t_\mathrm{avg}}$ that match the indicated SFR. 
            To avoid this issue, we determine best-fitting \tmin/ for \textit{ensembles of observation times} in windows of size $T_\mathrm{w} =500$ Myr (longer that the typical short time-scale variations during bursty phases).
            The increased accuracy in \tmin/ resulting from introducing the time window comes at the cost of decreased ability to localise \tmin/ in time within the window.
            We assign best-fitting values of \tmin/ to the end of time windows (rather than the middle, for example).
            This ensures that \tmin/ is computed using only SFRs that are causally connected to the observation time $t$.   
            We quantify how our results depend the exact choice of $T_\mathrm{w}$ in Appendix \ref{a:window} and find that they are stable for our fiducial choice of $T_\mathrm{w} =500$ Myr.
            
            To find the best-fitting \tmin/, we define another root mean square error which is a function of $t_\mathrm{avg}$:
            \begin{equation}
                \label{e:rmse}
                \mathrm{RMSE}(t_\mathrm{avg}) = 
                \sqrt{\frac{\sum_i \left(\log_{10}\sfrind/^i -\log_{10}\langle\sfrtrue/\rangle_{t_\mathrm{avg}}^i\right)^2}{n}},
            \end{equation}
            where $i$ is a sum over time samples spaced by 1 Myr covering the time window of total width $T_\mathrm{w}$. 
            This corresponds to the scatter with respect to the one-to-one line in log-space in the middle row of Figure \ref{f:scatter_summary}.  
            Within a given time window, we repeat this process and recompute SFHs of $\langle \sfrtrue/\rangle_{t_\mathrm{avg}}$ varying $1 \leq t_\mathrm{avg} \leq 1000$ Myr in steps of 1 Myr (see the bottom row of Figure \ref{f:scatter_summary}). 
            Lastly, we define the global minimum of this curve as \tmin/.
            We repeat this procedure in ${\sim}13000$ maximally overlapping windows (i.e. overlapping by 499 Myr) in order to construct histories of \tmin/ over cosmic time.
            We omit the first 1 Gyr of the simulation to exclude transient effects related to the initial formation of the dark matter halo and galaxy.

\section{Results}
    \label{s:results}
\subsection{Best-fitting time-scales}
\label{s:best_timescales}
        Figure \ref{f:scatter_summary} summarises our key results for two $T_\mathrm{w}=500$ Myr windows in \texttt{m12i}, a representative simulation of a Milky Way-mass halo. 
        The left column focuses on high-redshift period, when the star formation is bursty, while the right column focuses on a low-redshift period when the SFR is relatively time-steady.
        The top row shows the SFHs during these periods.
        At early times (left panel) the SFR changes by a factor of $> 10$ on time-scales of ${\sim} 30-50$ Myr whereas at late times (right panel) the SFR changes only by ${\lesssim} 10\%$.
        
        The middle-row panels show scatter plots of the true SFRs averaged over the best-fitting averaging time-scale, \tmin/, vs. the indicated SFRs for FUV 
        and \halpha/ for the times corresponding to the right edge of the top-row panels. 
        We also plot indicated SFHs for the often-quoted time-scales associated with each indicator (100 Myr and 10 Myr), rather than \tmin/, for comparison in gray. 
        The points for the 100 Myr and 10 Myr time-scales are offset vertically for clarity (by factors indicated on the panels). 
        Contours containing 50\% of points in each case are shown in solid black. 
        We see that the scatter for \tmin/ is substantially smaller than the scatter for the reference 100 Myr and 10 Myr time-scales. 
        
        The bottom row of Figure \ref{f:scatter_summary} shows the RMSE as a function of $t_\mathrm{avg}$, defined in Equation (\ref{e:rmse}), normalised by the RMSE at \tmin/ (the RMSE minimum). 
        We see little difference between the \tmin/ for \halpha/ when comparing the high- and low-redshift results (5 Myr vs. 4 Myr).
        This is because \halpha/ probes time-scales over which the SFR does not change drastically even in the bursty regime. 
        Thus, \halpha/ tracks the true, instantaneous SFR fairly accurately in both the bursty and steady regimes.
        
        FUV, on the other hand, shows more dependence on the burstiness of SFR. 
        For the example shown in the right column of Figure \ref{f:scatter_summary}, when SFR is time-steady, we find a best-fit \tmin/${\approx} 12$ Myr. 
        This is in contrast to the much longer best-fit \tmin/${\approx} 54$ Myr for the bursty period on the left experienced earlier by the same galaxy. 
        As we  discuss in more detail below, the best-fit \tmin/ can fluctuate strongly immediately following particularly extreme bursts of star formation, so the best-fit for this example is not necessarily representative. 
        
        A concrete example of how FUV responds to a strongly changing SFR is shown in the top left panel of \ref{f:scatter_summary} (see the vertical dashed line at $t=6.4$ Gyr). 
        The left panels of Figure \ref{f:sfr_maps} show renderings of the FUV and \halpha/ light from the galaxy at this time. 
        In this case, due to the strong drop in SFR shortly before the $t=6.4$ Gyr ``observation time,'' the observed FUV light was emitted from stars that formed ${\sim}50-100$ Myr in the past. 
        As a result, the FUV-indicated SFR is much larger than true instantaneous SFR, so that this indicated SFR is more representative of a time-average extending to when the stars dominating the observed emission formed.
        
        Next, we explore how \tmin/ varies as a function of cosmic time and galaxy mass. 
        Figure \ref{f:tmin_history} shows the evolution of \tmin/ for FUV and \halpha/ over cosmic time for \texttt{m10q}, \texttt{m11q}, and all four \texttt{m12} simulations listed in Table \ref{t:sim_table}. 
        For clarity, we show only results for a subset of the ${\sim}$13000 overlapping windows described in \S \ref{s:scatter}. 
        Specifically, in each panel we show results for 56 equally spaced windows, each overlapping by 250 Myr with its neighbors.
        
        We first focus on the two rightmost columns, corresponding to Milky Way-mass galaxies, which have dark matter haloes of mass $M_h \sim 10^{12}~\mathrm{M_\odot}$ at $z=0$ (the \texttt{m12} simulation plotted in red in the middle column is the same as in Figure \ref{f:scatter_summary}).
        These galaxies transition from bursty at high redshift to steady at later times. 
        Vertical dashed lines indicate the times when bursty star formation ends for each simulation, using the definition in \S \ref{s:SFHs}. 
        We see that, when the SFR is time-steady (to the right of the dashed lines), the best-fitting averaging time-scales for \halpha/ and FUV are relatively stable (especially for \halpha/) at $4-5$ Myr and $10-20$ Myr, respectively. 
        These ranges are indicated by the shaded regions in each row. 
        As we found for the example in Figure \ref{f:scatter_summary}, the best-fit \tmin/ for \halpha/ is, in most cases, nearly the same even at high redshift when SFRs are highly variable. 
        However, the best-fit \tmin/ for FUV fluctuates strongly in time, by factors up to $>10$ (i.e., up to values \tmin/$>100$ Myr). 
        Thus, during bursty periods there is a large uncertainty in the actual time-scales probed by FUV-indicated SFRs.

\begin{figure}
    \centering
    \includegraphics[width=\linewidth]{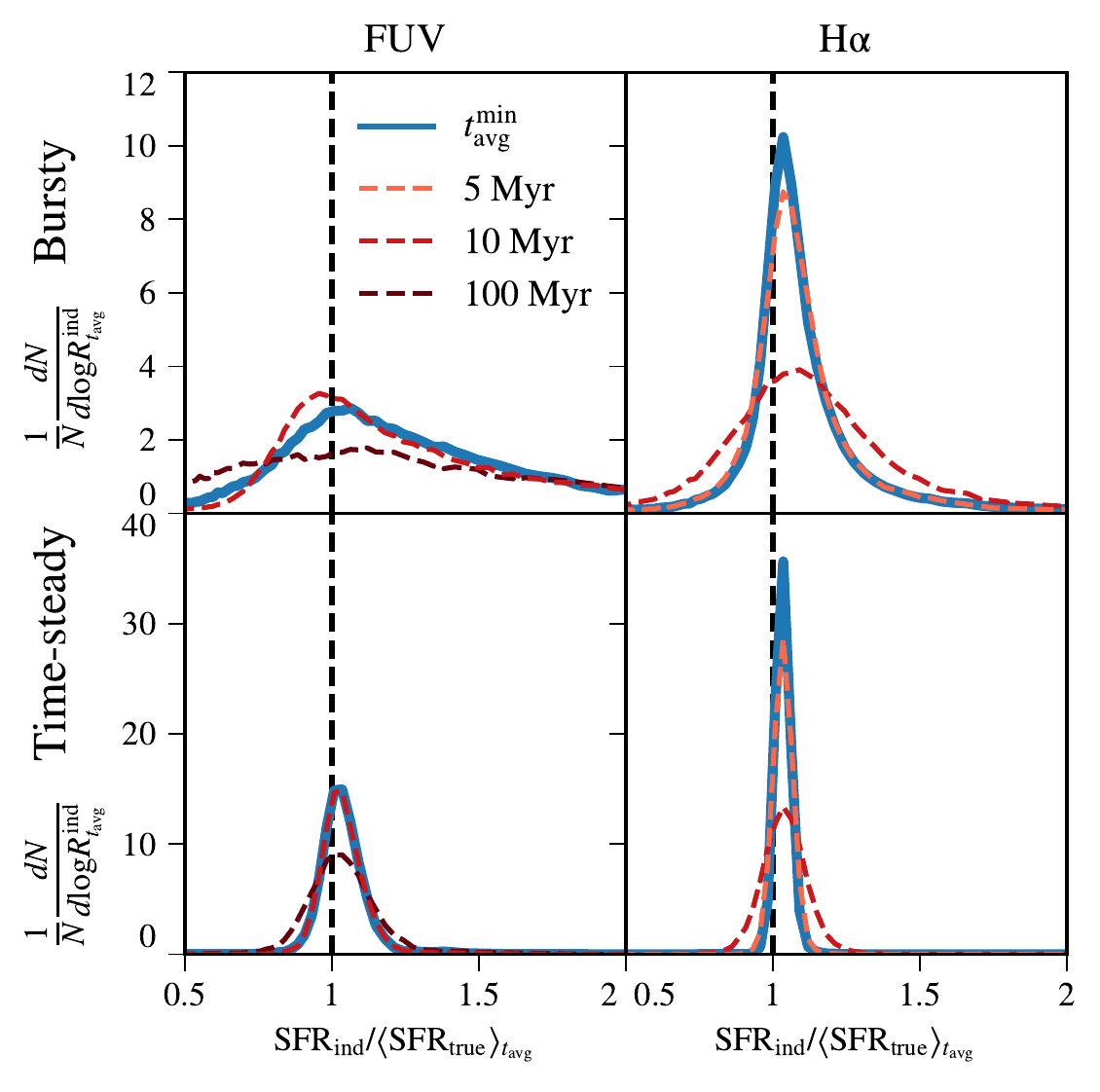}
    \caption{\label{f:ratio_scatter}
        Distributions of the indicated-to-time-averaged-true-SFR ratio, $R_{t_\mathrm{avg}}^\mathrm{ind} \equiv \mathrm{SFR_{ind}}/\langle \sfrtrue/ \rangle_{t_\mathrm{avg}}$, for different averaging timescales, $t_\mathrm{avg}$, and SFR indicators (FUV, left column; \halpha/, right column).
        SFHs from each simulation listed in Table \ref{t:sim_table} are partitioned into bursty (top row) and time-steady (bottom row) regimes and then combined to form each distribution plotted. 
        Distributions using \tmin/ are plotted in blue while different shades of red identify different fixed $t_\mathrm{avg}$ used to average the true SFR.  
        Vertical dashed lines are plotted at a ratio of unity, while the standard derivations of these distributions in dex are compiled in Table \ref{t:sigma_table}. 
        The scatter of the $\log_{10}{R_{t_\mathrm{avg}}^\mathrm{ind}}$ distribution, peaked near unity, is generally smallest for \tmin/, though similar to the scatter for the fixed 10 Myr and 5 Myr time-scales for FUV and \halpha/, respectively.
        }
\end{figure}

        The left column of Figure \ref{f:tmin_history} shows the results for dwarf galaxies in haloes of $z=0$ mass $M_h\sim 10^{10}$ M$_\odot$ and ${\sim} 10^{11}$ M$_\odot$ (\texttt{m10q} and \texttt{m11q}). 
        The SFHs for these dwarf galaxies are bursty at all times. 
        The more extreme burstiness for these dwarf galaxies (see the SFHs in Figure \ref{f:cosmic_SFH}) leads to more frequent excursions to very long values of \tmin/.
        For \texttt{m10q}, the figure shows that values of $\tmin/\gg 100$ Myr (up to ${\sim} 1$ Gyr) are found at some early times ($z \geq$ 2).
        While it is beyond the scope of this paper to directly compare the simulated SFHs presented here to those observed in the real universe, we note that the bursty SFHs in these dwarf galaxies may be consistent with the low quiescent fraction of observed dwarf galaxies at low redshift \citep[e.g.,][]{Geha2012, Dickey2020}. 
        
        The above results are illustrative of the main results of our analysis: 1) when the SFH is time-steady, \tmin/ is relatively short for both \halpha/ and FUV (${\sim} 5$ and ${\sim} 10$ Myr respectively); and 2) when the SFH is bursty, \tmin/ can be much longer, especially for FUV due to its greater sensitivity to longer time-scales. 
        As noted in \S \ref{sec:this_paper}, however, different simulations (with different models for star formation and stellar feedback) predict different SFR variability distributions, so the detailed quantitative results in this section (e.g., the range of \tmin/ values in bursty periods) are expected to differ for different simulations. 
        In steady periods, though, the best-fit time-scales do not depend on the detail of the simulations.
        
\begin{table}
    \setlength{\tabcolsep}{5pt}
    \renewcommand{\arraystretch}{1.5}
    \caption{\label{t:sigma_table} Intrinsic$^\mathrm{a}$ 1-$\sigma$ scatter in $\log_{10}{(\mathrm{SFR_{ind}}/\langle \sfrtrue/ \rangle_{t_\mathrm{avg}})}$ for different averaging time-scales in the bursty and time-steady star formation phases}
    \centering
        \begin{tabular}{@{}l@{} cllcc @{\hspace{-4pt}}ccccc@{}}
           &&&FUV  &&&&&\halpha/& \\[-2pt] \cline{3-5} \cline{8-10}
          SFR phase & & \tmin/ & 10 Myr & 100 Myr &&&\tmin/ & 5 Myr & 10 Myr\\ \cline{1-1}\cline{3-5} \cline{8-10} 
          Bursty & & 0.143$^\mathrm{b}$ & 0.120$^\mathrm{b}$ & 0.242 &&& 0.035 & 0.041 & 0.094 \\ 
          Time-steady & & 0.025 & 0.026 & 0.042 &&& 0.011 & 0.014 & 0.030 \\  \hline
        \end{tabular}\\
        \begin{flushleft}
        $^\mathrm{a}$ This neglects uncertainties from observational effects, stellar population synthesis modeling, dust obscuration, etc.\\
        $^\mathrm{b}$ Note that in this case the 10 Myr average produces a scatter smaller than (but similar to) the \tmin/ average. 
        This is possible because the minimization procedure used to obtain \tmin/ is different from minimizing the width of the $R_{t_\mathrm{avg}}^\mathrm{ind}$ distribution (see \S \ref{s:scatter}).
        As a result, the time-scale that minimizes the RMSE in a 500 Myr window is not guaranteed to also minimize the 1$-\sigma$ scatter of the corresponding log-ratio distribution (Figure \ref{f:ratio_scatter}).
        Moreover, stochasticity in the SFH, especially in the bursty regime, can introduce noise in the numerical estimates of scatter.
        \end{flushleft}
    \end{table}

\subsection{Theoretical uncertainty due to unknown SFH}
\label{s:indicated_errors}
        In the previous section, we quantified the best-fitting time-scales for FUV and \halpha/ indicators in a statistical sense. 
        Even when the relevant time-scale is known, an important question is: how well does the indicated SFR match the true SFR boxcar-averaged over \tmin/ when a measurement is made? 
        In general, $\langle {\rm SFR}_{\rm true} \rangle_{\tmin/} \neq {\rm SFR}_{\rm ind}$ because, depending on the SFH history, there may be \emph{no} value of $t_{\rm avg}$ for which the indicated SFR exactly matches the boxcar average over the preceding $t_{\rm avg}$. 
        This introduces a minimum uncertainty (absent other information) in any measurement due to the unknown shape of the recent SFH, on top of the modeling uncertainties associated with the assumed stellar population synthesis model (including the IMF) and dust obscuration.
        
        Figure \ref{f:ratio_scatter} quantifies this uncertainty by plotting distributions of the indicated-to-time-averaged-true-SFR ratio, $R_{t_\mathrm{avg}}^\mathrm{ind} \equiv \mathrm{SFR_{ind}}/\langle \sfrtrue/ \rangle_{t_\mathrm{avg}}$, for different averaging time-scales $t_\mathrm{avg}$.
        To produce these distributions, we first partition the simulations listed in Table \ref{t:sim_table} into their bursty and time-steady phases (the two rows in the figure). 
        For each of these phases, we consider the FUV and \halpha/ indicators separately (the two columns). 
        Then, for each panel, we evaluate the ratio $R_{t_\mathrm{avg}}^\mathrm{ind}$ for $t_{\rm avg}=\tmin/,$ 100 Myr, 10 Myr, and 5 Myr at 1 Myr intervals (at each time, $\tmin/$ is obtained by minimizing over a time window of total width $T_{\rm w}=500$ Myr, as described in \S \ref{s:scatter}). 
        The distributions are shown for the results from all the simulations combined.
        
        We see that, in almost all panels, \tmin/ produces the smallest scatter around $R_{t_\mathrm{avg}}^\mathrm{ind} \approx 1$, as expected since \tmin/ is defined as the best-fitting time-scale in each case. 
        The $R_{t_\mathrm{avg}}^\mathrm{ind}$ distributions also confirm the results from the previous section that $\tmin/ \sim 10$ Myr for FUV and $\tmin/ \sim 5$ Myr for \halpha/ (see Figure \ref{f:tmin_history}), since the distributions are similar when using either $\tmin/$ or the corresponding constant averaging time-scale.
        On the other hand, the scatters for $t_{\rm avg}=100$ Myr and $t_\mathrm{avg}=10$ Myr are significantly larger in all panels, indicating again that these time-scales are typically much longer than those effectively probed by either FUV or \halpha/.
        
        To further quantify the $R_{t_\mathrm{avg}}^\mathrm{ind}$ distributions, we compute the standard deviations of $\log_{10} R_{t_\mathrm{avg}}^\mathrm{ind}$ for different cases and list the results in Table \ref{t:sigma_table}.  
        For comparison, the typical observational uncertainty in SFRs measured through monochromatic indicators, combinations of monochromatic indicators, or SED modeling is ${\sim}0.1-0.3$ dex \citep[e.g.][]{Hao2011,Koyama2015,Iyer2019}. 
        Thus, the 0.12-0.14 dex intrinsic scatter that we predict for FUV in the bursty regime for $\tmin/ \approx 10$ Myr, due solely to the ``unknown'' preceding SFH, can be a significant additional source of uncertainty. 
        On the hand, the intrinsic scatters ${\approx} 0.01-0.04$ dex predicted for \halpha/ (in the steady and bursty regimes) for $\tmin/ \approx 5$ Myr imply that this intrinsic uncertainty is usually subdominant.

\section{Discussion}
    \label{s:discussion}
\begin{figure}
    \centering
    \includegraphics[width=\linewidth]{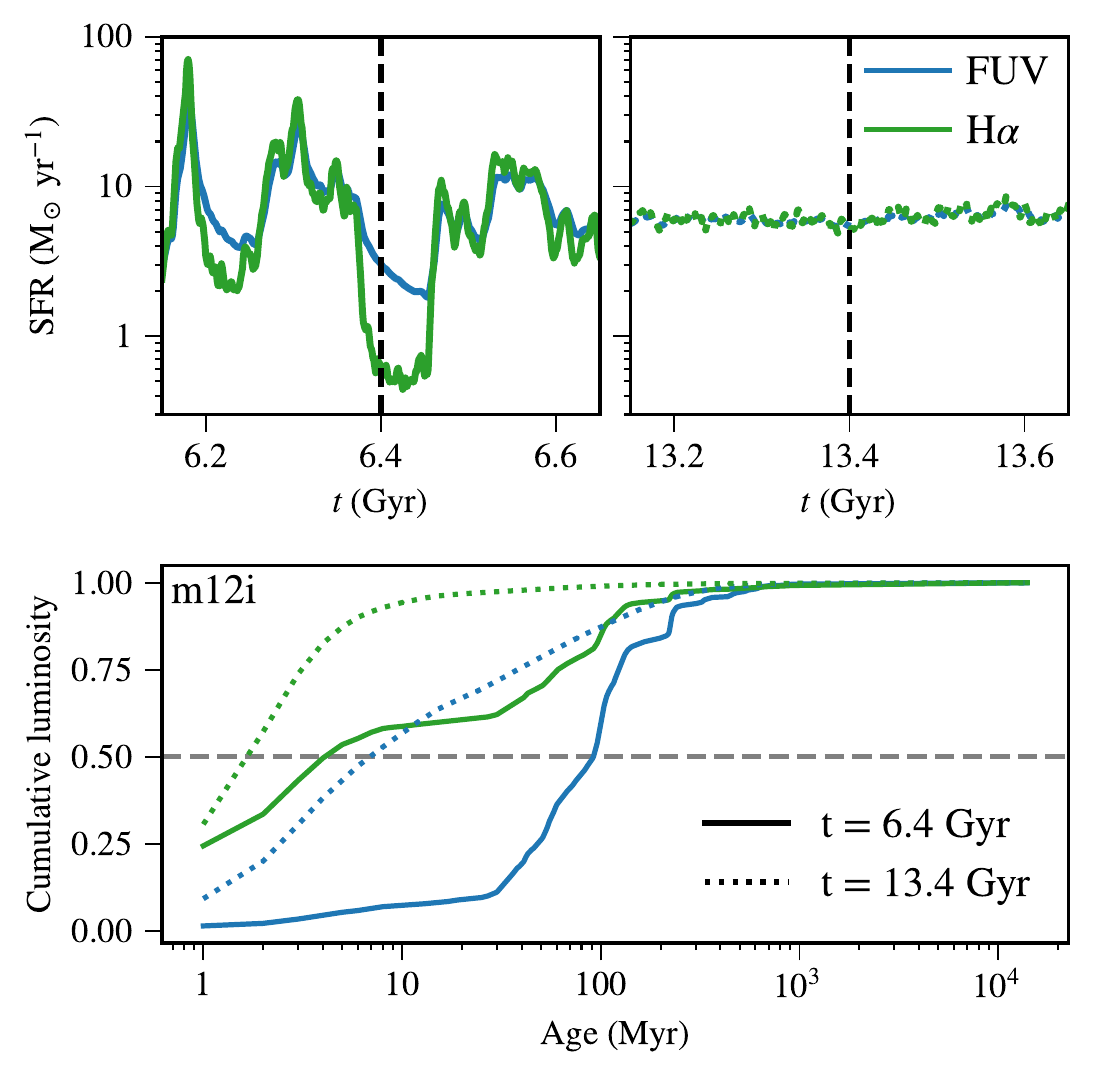}
    \caption{\label{f:percentile_def} 
    \textit{Top:} SFHs near cosmic-times $t=6.4$ Gyr ($z=0.9$ and ${\approx} 50$ Myr after a burst of star formation; left) and $t=13.4$ Gyr ($z=0.03$ and long after the SFR has settled to a steady state; right) as indicated by FUV (blue) and \halpha/ (green) for \texttt{m12i}.
    \textit{Bottom:} Cumulative luminosity in FUV and \halpha/ as a function of stellar age, at $t= 6.4$ Gyr (solid curves) and $t=13.4$ Gyr (dotted curves).
    A horizontal dashed line is plotted at $50\%$ to identify the intersection with each cumulative luminosity curve, which we define as \tfifty/ (see \S \ref{s:tfifty}).
    For FUV \tfifty/ is strongly dependent on the recent SFH and is larger when the SFR is bursty compared to when it is time-steady. 
    For \halpha/, \tfifty/ also varies depending on the recent SFH but by a smaller factor (${\sim} 2\times$ in the example shown).}
\end{figure} 
\begin{figure}
    \centering
    \includegraphics[width=\linewidth]{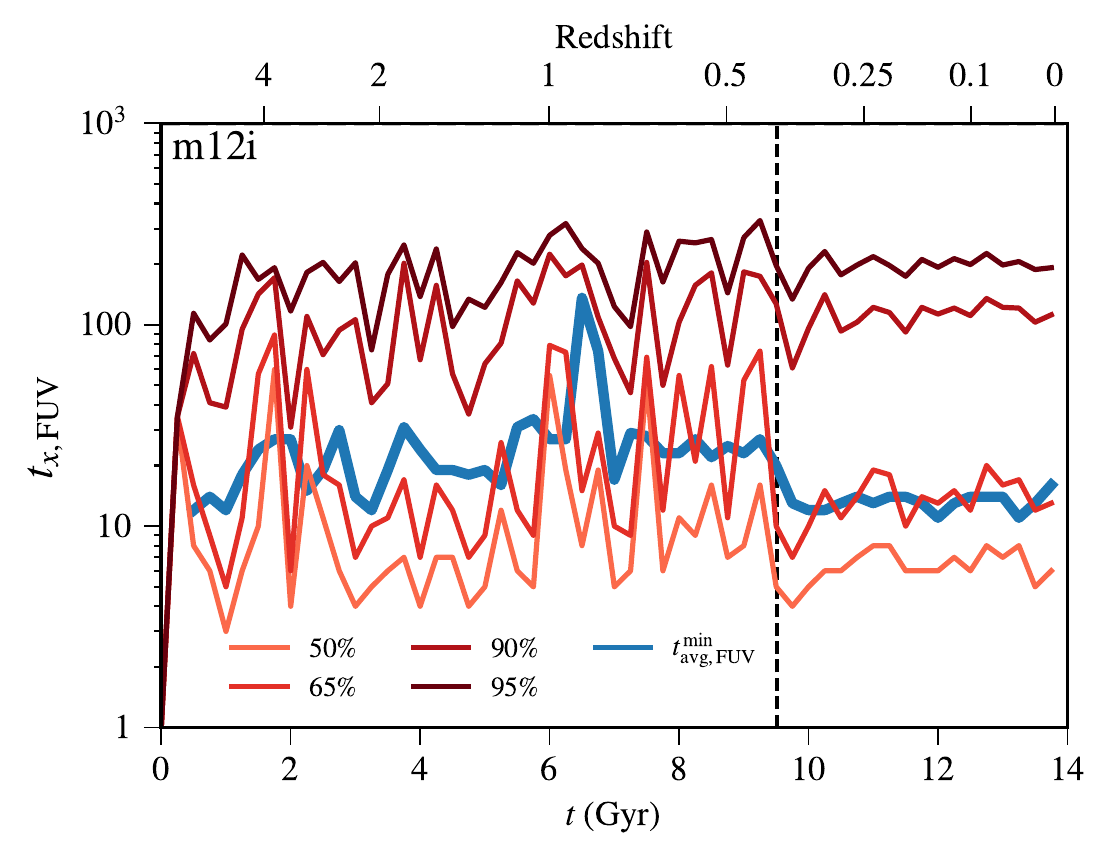}
    \caption{\label{f:percentiles} FUV light-weighted percentile stellar ages, $t_{x}$, for different percentile values ($x=50\%, 65\%, 90\%$, and $95\%$) as a function of cosmic time in \texttt{m12i}. 
    The best-fitting boxcar-average equivalent time-scale, \tmin/, is also shown for comparison (solid blue), along with the end of bursty star formation (vertical dashed line).
    Both \tmin/ and $t_x$ fluctuate much more strongly during the bursty phase. 
    The fluctuations can cause the different $t_x$ curves to reach values well above the quasi-steady values predicted at late times, especially, for the smaller percentiles, which are more sensitive to short time-scale SFR fluctuations.
    }
\end{figure}

    \subsection{Range of time-scales probed}
    \label{s:tfifty}
    So far, we have characterized the time-scales probed by SFR indicators by \tmin/, which is the equivalent width of a boxcar-average response kernel.
    In reality, observational SFR indicators probe light emitted by stars with a range of different ages. 
    This is especially true for FUV light, whose emission persists long after the \halpha/ emission powered by ionizing radiation becomes negligible (e.g., left panel of Figure \ref{f:indicators}). 
    Another way to characterize the time-scales probed by an indicator is to partition the total luminosity observed at a time $t$ by the age of the stars contributing to the light.
    At any given cosmic time $t$, we define \tfifty/ as the stellar age below which the combined luminosity in the band of interest is $50\%$ of the total luminosity, i.e. the light-weighted median stellar age.
    Similarly, we define $t_{x}$ for $x=65\%$, $90\%$ and $95\%$.
    These time-scales reflect the extent to which older stars contribute to the observed luminosity compared to younger stars.
        
    Figure \ref{f:percentile_def} illustrates how the total observed luminosity is contributed to by stars of different ages for two different cosmic times in \texttt{m12i}, one during a period in the bursty phase and one in the time-steady phase. 
    Because the time-scale of \halpha/ is short compared to most SFR variations, it tracks the true SFR with relatively little time delay in either the bursty phase or the steady phase ($\tfifty/ \sim 2-4$ Myr). 
    On the other hand, FUV exhibits a much larger difference between the two phases of star formation, with \tfifty/ fluctuating from ${\sim} 5$ Myr to ${\sim}$ 50 Myr in the bursty phase, before stabilizing in the time-steady phase.

    Figure \ref{f:percentiles} demonstrates in more detail how the stellar age distribution of observed FUV photons fluctuates by comparing \tmin/ to different $t_{x}$ for \texttt{m12i} as a function of cosmic time.
    Both \tmin/ and $t_x$ fluctuate much more strongly during the bursty phase (left of the vertical dashed black line) than in the more steady SFR phase (right of the vertical line). 
    The fluctuations can cause the different $t_x$ curves to reach values well above the quasi-steady values predicted at late times. 
    This is especially so for the smaller percentiles, which are more sensitive to short time-scale SFR fluctuations. 
    As expected from previous studies, the asymptotic value of $t_{90\%}\sim 100$ Myr is consistent with the time-scale sometimes associated with FUV, defined such that FUV reach 90\% of its asymptotic luminosity for a constant SFH (see \S \ref{s:measuring_observationally}). 
    In general, the distribution and time evolution of $t_{x\%}$ values depend on the details of the SFH, and so can depend on galaxy mass and redshift.
    However, for all simulations that achieve time-steady SFRs at late-times, the $t_{x\%}$ that best matches the late-time \tmin/ is $t_{65\%}$.

\subsection{\halpha/-to-FUV ratio as a gauge of SFR burstiness and its relation to \tmin/ and \tfifty/}
    \label{s:sfr_ratio}
\begin{figure}
    \centering
    \includegraphics[width=\linewidth]{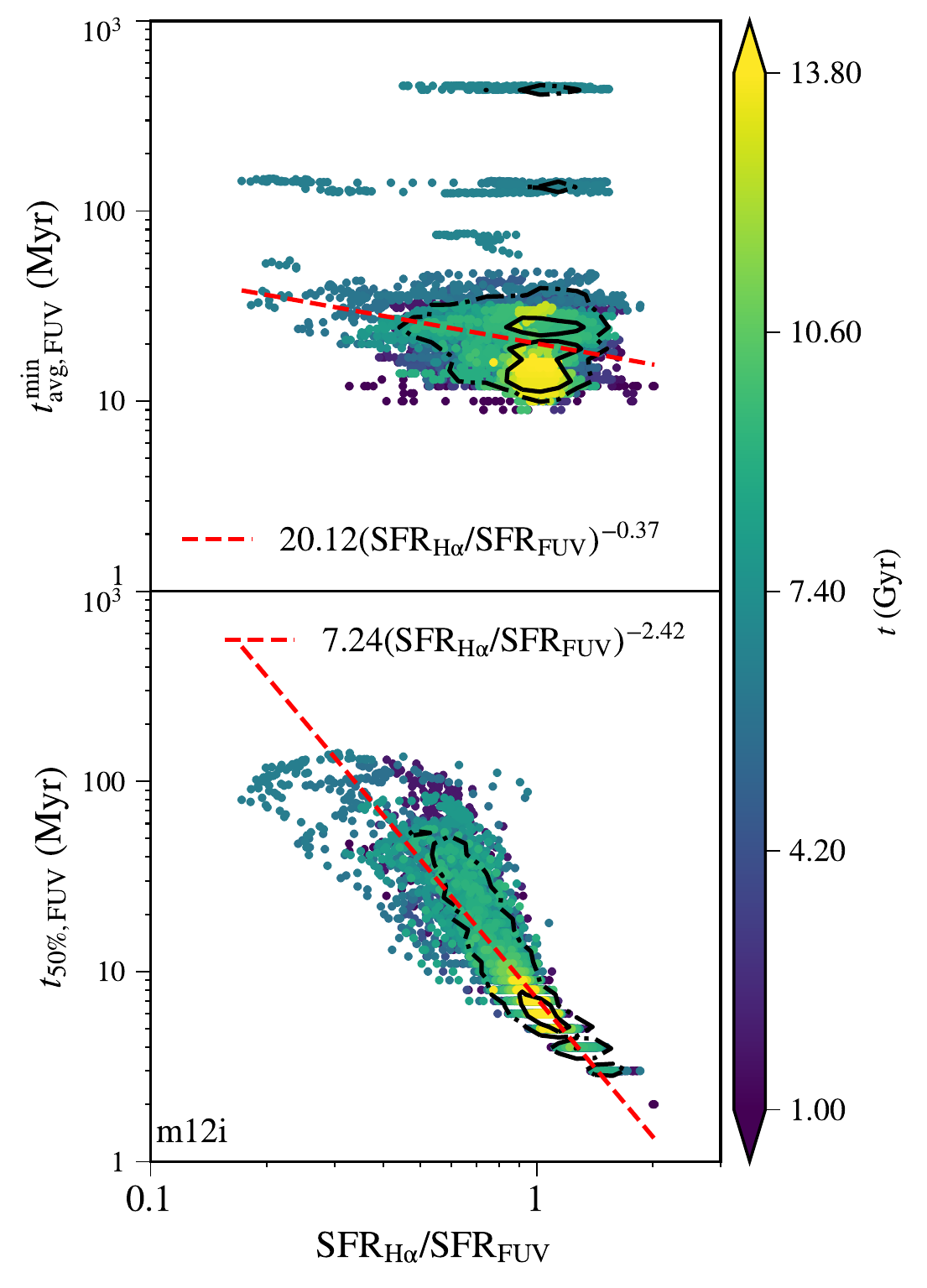}
    \caption{\label{f:indicator_ratio} Scatter plots of the best-fitting boxcar-average equivalent time-scale, \tmin/ (top) and the light-weighted median stellar age, $t_{50\%}$, for FUV vs. the instantaneous \halpha/-to-FUV ratio in \texttt{m12i}. 
    Each dot represents the value at one point of the full SFH, sampled at 1 Myr intervals (in each case, the first 1 Gyr is omitted to avoid edge effects near $t=0$). 
    50$^\mathrm{th}$ and 90$^\mathrm{th}$ percentile contours are plotted in solid and dot-dashed black respectively. 
    A least-squares best-fitting line is plotted in dashed red, with parameters indicated in each panel. 
    For \tmin/, there is almost no correlation with the SFR ratio because \tmin/ is computed by minimizing over a long time window $T_\mathrm{w}=500$ Myr (see \S \ref{s:scatter}). 
    As a result, \tmin/ is a poor measure of the recent SFH probed by the instantaneous \halpha/-to-FUV ratio in bursty phases. 
    On the other hand, like the \halpha/-to-FUV ratio, \tfifty/ is directly sensitive to the recent SFH and is thus much more tightly anti-correlated with ${\rm SFR}_{\halpha/}/{\rm SFR}_{\rm FUV}$.
    }
\end{figure}

        Since FUV and \halpha/ probe the SFR averaged over different time-scales, the ratio of FUV to \halpha/ can be used as a probe the SFR variability \citep[e.g.,][]{Fumagalli2011, Weisz2012,Broussard2019}. 
        As the best-fit time-scales \tmin/ and \tfifty/ are also sensitive to changes in the SFR, we investigate whether the observed \halpha/-to-FUV ratio can be used to infer more accurate (i.e. more applicable to the specific time corresponding to the observation)  time-scales than the ones listed above that are implied statistically by the FIRE-2 simulated SFHs.
        
        Figure \ref{f:indicator_ratio} shows a scatter plot of the \halpha/-to-FUV ratio vs. \tmin/ (top panel) and \tfifty/ (bottom panel),  for \texttt{m12i} and cosmic times $1 < t < 13.8$ Gyr spaced by 1 Myr (this includes both times when the SFR is bursty and when it is steady). 
        This figure focuses on FUV time-scales because, owing to their longer values, they are more sensitive to variations in the SFH than \halpha/. 
        For \tmin/, we see almost no correlation with the SFR ratio.
        This is because, as we defined it, \tmin/ is computed by minimizing over a long time window $T_\mathrm{w}=500$ Myr. 
        As a result, since the SFR (and consequently the \halpha/-to-FUV ratio) fluctuates on time scales smaller than $T_{\rm w}$, during bursty phases \tmin/ is a poor measure of the very recent (or ``local'') SFH, which is the portion of the SFH probed by the instantaneous \halpha/-to-FUV ratio. 
        Thus, within a single window while \tmin/ remains effectively constant the \halpha/-to-FUV ratio takes many values (both less than and larger than unity) tracing the horizontal lines observed in the top panel of Figure \ref{f:indicator_ratio}, and washing out any correlation between the two.
        
        On the other hand, like the \halpha/-to-FUV ratio, \tfifty/ is directly sensitive to the SFH at times immediately preceding the time of observation.
        This results in a much tighter anti-correlation between \tfifty/ and the \halpha/-to-FUV ratio, visible in the bottom panel of the figure.
        The slope of relationship is negative because a smaller \halpha/-to-FUV ratio implies that (on average) the SFR has decreased from the earlier times probed by FUV light to the more recent times probed by \halpha/ light. 
        This implies that the stellar age of the stars below which 50\% of the observed FUV luminosity was produced, i.e. $t_{\rm 50\%,FUV}$, is larger (i.e. formed during a recent burst of star formation). 
        Quantitatively, the best-fitting the slope of the power-law relation plotted in the bottom panel of the figure is $-2.42$ for \texttt{m12i}, though it can vary with galaxy mass and redshift as the statistics of SFR variations evolve.
        In addition to splitting the SFH into two distinct bursty and time-steady phases, and quantifying the best-fitting time-scales in each regime as we have done above, 
         a Fourier analysis of the SFH can quantify the variability on different time-scales \citep{Tacchella2016,Caplar2019,Iyer2020,Wang2020}.
         \cite{Caplar2019} point out that, in their formalism, both the slope of the PSD on short time-scales and the ``decorrelation time-scale,'' $\tau_\mathrm{break}$ (the time-scale after which the PSD resembles white noise) are accessible from observations of the \halpha/-to-FUV ratio. 
         \cite{Iyer2020} directly quantifies the PSD of the ``true'' SFHs in the FIRE-2 simulations. 
        
        In future work, it would be interesting to expand our analysis to connect more directly to PSD statistics, as well as to more systematically quantify the time-scales probed by SFR indicators as a function of the observed \halpha/-to-FUV ratio.
        
        \subsection{Caveats and possible extensions}
        \label{sec:caveats}
        \subsubsection{Dust attenuation}
        \label{sec:dust_attenuation}
            In this work, we have neglected attenuation of UV light by dust \citep[e.g.,][]{Calzetti2000, Reddy2016}. 
            This assumption is reasonable for dwarf galaxies that contain little dust \citep[e.g.,][]{Broussard2019}, but attenuation by dust is in general critical to model to obtain correctly normalized SFR measurements. 
            Since attenuation by dust depends on wavelength, it can also affect the ratio of different SFR indicators, such as the \halpha/-to-FUV ratio discussed above. 
            Dust could modify the time-scales probed by an SFR indicator if the attenuation that it produces is time-dependent and correlated with the light sources \citep{Koyama2019, Salim2020}. 
            For example, if dust initially obscures star-forming regions but is cleared by stellar feedback after some time, this could modify the effective response function of the indicator. 
            Properly modeling this kind of effect requires accounting for the exact spatial distribution and time-dependence of both the dust and light sources in galaxies, along with a detailed treatment of radiative transfer. 
            
            Zoom-in simulations such as the FIRE-2 simulations analyzed in this paper can be combined with dust radiative transfer codes \citep[e.g.,][]{Jonsson2006, Camps2015, Narayanan2020} to model these effects. 
            This would also make it possible to study SFR indicators resulting from the processing of direct stellar radiation by dust, such as indicators based on IR radiation.
            A preliminary analysis of our simulations including dust radiative transfer suggests that, especially during bursty phases, 
            there are substantial time-varying effects from dust attenuation.
            These effects are generally larger on FUV than on \halpha/, and we find that the magnitude of dust attenuation is strongly correlated with the true SFR, such that the emergent FUV flux can be lower when SFR is higher.
            
            In this paper, we have assumed that observational techniques can correct for dust attenuation sufficiently accurately to recover intrinsic FUV and H$\alpha$~fluxes that can be compared to our idealized analysis, but in the future it will be important to more explicitly forward-model the radiative transfer to assess when this can be done robustly.

            \subsubsection{IMF sampling and variations}
            
            Another possible extension of this work would be to include the effects of stochastic sampling of the IMF, which could have interesting effects especially in dwarf galaxies \cite[e.g.,][]{daSilva2014}.
            Stochastic sampling of the IMF can have two different kinds of effects related to SFR burstiness. 
            The first is an apparent signature of SFR variability in observed light. 
            For example, even if a galaxy has a perfectly steady SFH, at low SFR its H$\alpha$ emission will fluctuate due to the fact that massive OB stars, which dominate the production of ionizing photons, form only rarely. 
            In the limit of very low SFRs, these massive stars can be so rare that their number fluctuates significantly from time to time. 
            The consequent, unavoidable light curve fluctuations introduce biases in SFRs inferred using indicators calibrated under the assumption of a steady luminosity (see \S \ref{s:indicators}). 
            The light curve variability will also cause fluctuations in the H$\alpha$-to-FUV ratio and could give the incorrect impression that the underlying SFH is fluctuating more than it really is. 
            
            The other effect of stochastic IMF sampling is dynamical and arises through the stellar feedback. 
            In the FIRE-2 simulations analyzed in this paper, the feedback processes are IMF averaged, meaning e.g. that each star particle has a radiative luminosity and supernova rate proportional to the rates for a stellar population that fully samples the IMF. 
            For dwarf galaxies for which the mass of individual star particles can be $\approx 250$ M$_{\odot}$ (our m10q run) or lower, some star particles should in reality contain no O star, which would significantly modify the feedback from those star individual particles. 
            Although such dynamical effects may be significant for some of the lowest-mass dwarfs and/or for certain subgrid models \citep[e.g.,][]{Applebaum2020}, explicit comparisons with re-simulations that do include the effects of stochastic sampling on feedback suggest that stochastic IMF sampling does not substantially modify the SFH statistics for the FIRE-2 simulations analyzed in this paper \cite[][]{Su2018,Wheeler2019}.

            Additionally, if the IMF varies with environment or galaxy properties \citep[e.g.,][]{vanDokkum2010}, the luminosity response functions for the SFR indicators would also vary, potentially altering the time-scales of SFR indicators. 
            \cite{Broussard2019} showed that the ratio of the near-ultraviolet (NUV, ${\sim}$2750 \AA)-to-\halpha/, $\eta$, is an additional indicator of bursty star formation on time-scales similar those identified for FUV here.
            They find that when the slope of the high-mass end of the IMF is steepened, the effective the time-scale probed by the $\eta$ ratio decreases. 
            This is because emission in both \halpha/ and the NUV becomes even more dominated by short-lived O and B stars. 
            IMF variations would also introduce dynamical feedback effects which could affect the SFHs predicted by simulations.

\section{Conclusions} 
    \label{s:conclusion}
    
    In this paper we quantify the time-scales probed by the FUV and \halpha/ indicators commonly used to measure star formation rates by using star formation histories from FIRE-2 cosmological zoom-in simulations \citep{Hopkins2018}. 
    A key goal of our study is to investigate the effects of realistic SFHs, including secular trends as well as short-term variability (``burstiness''), on the time-scales probed by standard indicators calibrated to constant SFHs \citep[e.g.,][]{Kennicutt2012}. 
    The SFHs for Milky Way-mass galaxies in the FIRE simulations exhibit a transition from a highly time-variable, or bursty, phase at early times to a time-steady phase at late times. 
    Moreover, dwarf galaxies are bursty at all epochs (Figure \ref{f:cosmic_SFH}).
    Our analysis is on based mock, unattenuated FUV and \halpha/ light curves produced by combining SFHs from FIRE-2 simulations with the \texttt{BPASS} stellar population synthesis model, which includes the effects of binary stars \citep[][]{Eldridge2017}.
    
    We define \tmin/, the best-fitting boxcar-average equivalent time-scale, which is a statistical measure computed in 500 Myr-wide moving windows that cover representative samples of short-time-scale SFR fluctuations (see \S \ref{s:scatter}).
    We find that $\tmin/ \sim 5$ Myr for \halpha/, independent of whether the SFH is bursty or time-steady. 
    On the other hand, \tmin/ fluctuates more during the bursty phase of star formation, from $\tmin/\sim10$ Myr to $>100$ Myr in cases following particularly extreme bursts of star formation. 
    However, when the SFR becomes time-steady, the FUV \tmin/ approaches ${\sim} 10$ Myr (Figure \ref{f:tmin_history}). 
    These best-fitting boxcar-average equivalent time-scales are notable because theoretical models often assume that FUV and \halpha/ observations probe SFRs boxcar-averaged over the much longer time-scales of 100 Myr and 10 Myr, respectively \citep[e.g.,][]{Orr2018, Hani2020}. 
    This difference arises because, for example, while stellar populations emit significant FUV for ages exceeding 100 Myr, the integrated UV emission is strongly weighted toward younger, massive stars. 
    
    In order to further characterize the distribution of stellar ages contributing to each SFR indicator, we also define \tfifty/ as the light-weighted median stellar age, and analogous quantities for different age percentiles. 
    Unlike \tmin/, \tfifty/ is a local measure of the recent SFH, in the sense that it is computed directly based on the SFH immediately preceding the time of interest (or observation) rather than being a statistical best fit over a large sampling window.
    We focused our analysis of the stellar age distribution on FUV (Figure \ref{f:percentiles}), which probes a wider age range. 
    For FUV, \tfifty/ fluctuates strongly, from ${\sim} 5$ Myr to ${\sim} 50$ Myr in the bursty phase for the example shown in the figure.
    As expected from previous studies, the asymptotic value of $t_{90\%}\sim 100$ Myr is consistent with the time-scale sometimes associated with FUV, defined such that FUV reach 90\% of its asymptotic luminosity for a constant SFH.
    The distribution and time evolution of $t_{x\%}$ values in general depend on the details of the SFH. 
    However, we find that for all simulations that achieve time-steady SFRs at late-times, the $t_{x\%}$ that best matches the late-time \tmin/ is $t_{65\%}$.
    
    We also investigated how the observed \halpha/-to-FUV ratio can tell us more about the time-scale probed by FUV light. 
    Since \halpha/ and FUV are sensitive to light from stars of different ages, the observed ratio is sensitive to the shape of the recent SFH. 
    In turn, the shape of the recent SFH affects $t_{50\%,{\rm FUV}}$. 
    Indeed, the simulations predict a strong anti-correlation between $t_{50\%,{\rm FUV}}$ and SFR$_{\halpha/}$/SFR$_{\rm FUV}$ (Figure \ref{f:indicator_ratio}). 
    This anti-correlation arises because, in periods following a burst of star formation, the FUV luminosity is dominated by older stars formed during the burst (corresponding to large $t_{50\%,{\rm FUV}}$) while the \halpha/ luminosity is dominated by younger stars. 
    Thus, following a burst after which the true SFR decays SFR$_{\halpha/}$/SFR$_{\rm FUV}$ decreases to ${<}1$ . 
    More generally, our results support the use of the \halpha/-to-FUV ratio as an observational probe of SFR variability \citep[e.g.,][]{Weisz2012}.
    
    Finally, we note that there are several avenues for extending the present study, which could lead to further important insights.
    These include the effects of dust attenuation and reprocessing of stellar light into IR indicators, stochastic sampling and possible environmental variations of the IMF, as well as the use of more detailed panchromatic data. 
    It is also important to reiterate that the detailed quantitative results we have presented apply to the set of FIRE-2 simulations we have analyzed. 
    Different galaxy formation simulations predict different SFR variability statistics \citep[][]{Iyer2020}, due e.g. to different models for star formation and stellar feedback. 
    For this reason, comparisons with observational statistics that quantitatively constrain the SFR variability in observed galaxy populations \citep[e.g.,][]{Sparre2017, Broussard2019, Caplar2019,Wang2020}, will play a key role, in combination with the methodology presented in this paper, in more precisely quantifying the time-scales probed by different indicators.

\section*{Acknowledgements}
This paper represents the effort of Jos\'e Antonio Flores Vel\'azquez, whose life was tragically taken on August 14, 2019 before he could see this project finished. 
Jos\'e was a beloved son, brother, friend, and colleague, with a warm smile and a big heart. Those who knew him were inspired by his unwavering optimism and passion for his work. 
On top of being a great person, he was also a remarkably talented astronomer, with great potential as exemplified by his drive to learn things by himself whilst also recognizing when to reach out for help. 
His refusal to sacrifice his identity also made, and continues to make, him a role model to folks who struggle to embody their culture in academic settings. 
The field of Astronomy lost a brilliant scientist and our goal is to share his work with the community. 

The authors thank the anonymous referee for their careful review of this manuscript. 
JAFV was supported by a National Science Foundation Graduate Research Fellowship Program under grant DGE-1839285.  
ABG was also supported by an NSF-GRFP under grant DGE-1842165  and was additionally supported by NSF grants DGE-0948017 and DGE-145000.
CAFG was supported by NSF through grants AST-1412836, AST-1517491, AST-1715216, and CAREER award AST1652522, by NASA through grants NNX15AB22G and 17-ATP17-0067, by STScI through grants HST-GO-14681.011, HST-GO-14268.022-A, and HST-AR-14293.001-A, and by a Cottrell Scholar Award from the Research Corporation for Science Advancement. 
JSB, AL, and FJM were supported by NSF grants AST-1910346 and AST-1518291. 
Support for JM is provided by the NSF (AST Award Number 1516374), and by the Harvard Institute for Theory and Computation, through their Visiting Scholars Program.
JS is supported as
a CIERA Fellows by the CIERA Postdoctoral Fellowship
Program (Center for Interdisciplinary Exploration and Research in Astrophysics, Northwestern University).
AW received support from NASA through ATP grant 80NSSC18K1097 and HST grants GO-14734, AR-15057, AR-15809, and GO-15902 from STScI; the Heising-Simons Foundation; and a Hellman Fellowship.
KE was supported by an NSF GRFP and a Hellman fellowship from UC Berkeley.
The Flatiron Institute is supported by the Simons Foundation. 
Numerical calculations were run on the Quest computing cluster at Northwestern University, the Wheeler computing cluster at Caltech, XSEDE allocations TG-AST130039, TG-AST120025 and TG-AST140023, Blue Waters PRAC allocation NSF.1713353, and NASA HEC allocations SMD16-7592, SMD-16-7561 and SMD-17-1204. 
We honour the invaluable labour of the maintenance and clerical staff at our institutions, whose contributions make our scientific discoveries a reality. 
This research was conducted on Potawatomi Indigenous land.

\section*{Data Availability}
The data supporting the plots within this article are available on reasonable request to the corresponding author. A public version of the GIZMO code is available at \url{http://www.tapir.caltech.edu/~phopkins/Site/GIZMO.html}. 
Additional data including simulation snapshots, initial conditions, and derived data products are available at \url{http://fire.northwestern.edu/data/}.

\appendix
\section{Dependence of our results on width $T_\mathrm{w}$ of the time window}
\label{a:window}
\begin{figure}
    \centering
    \includegraphics[width=\linewidth]{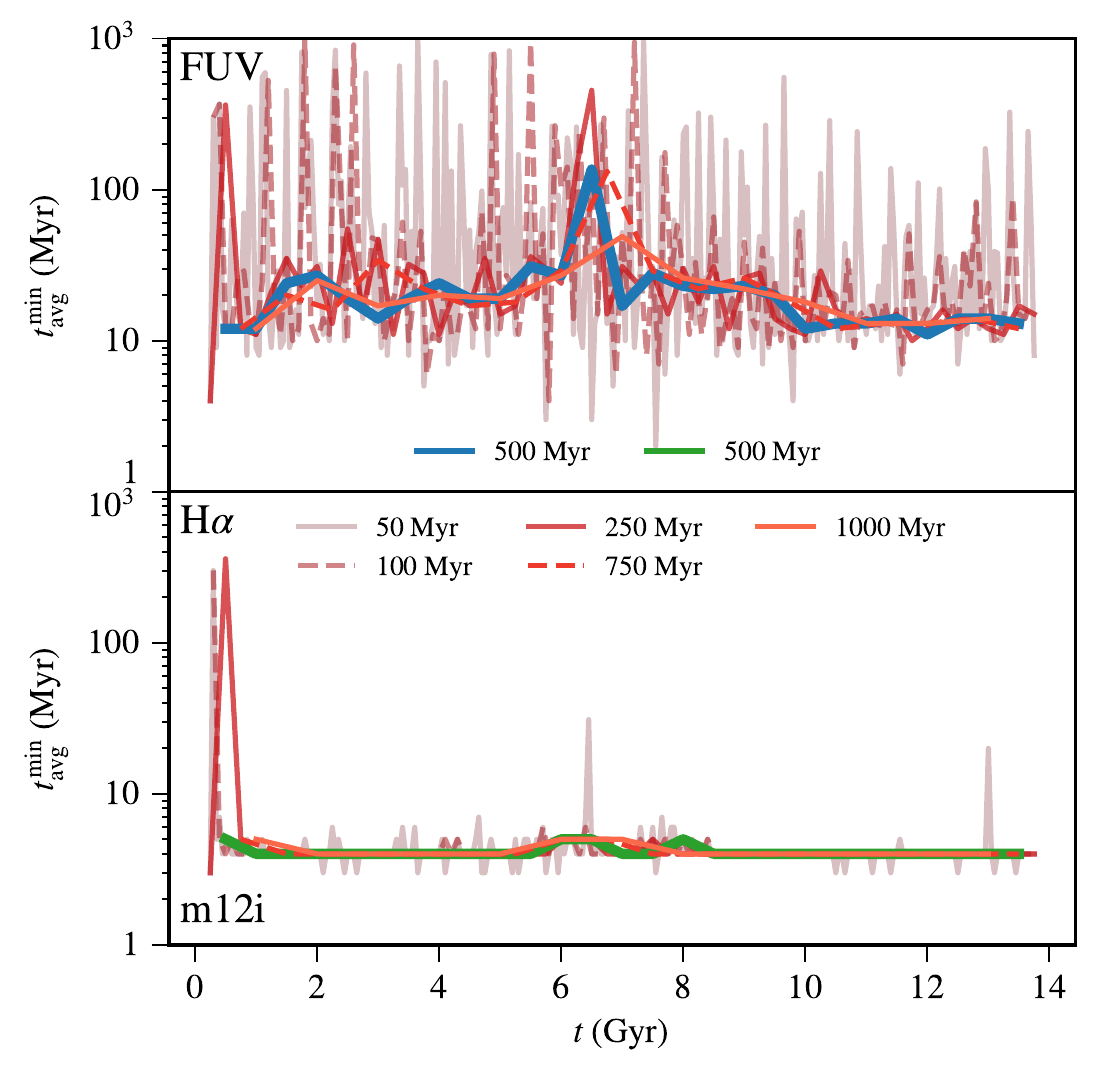}
    \caption{\label{f:window_test}
        History of \tmin/ over cosmic time for FUV (top) and \halpha/ (bottom) in non-overlapping windows for different choices of window width $T_\mathrm{w}$ (indicated by color) in \texttt{m12i}.
        For sufficiently large $T_\mathrm{w}$ (${\gtrsim} 250$ Myr for FUV and ${\gtrsim} 50$ Myr for \halpha/) \tmin/ is relatively insensitive to the choice of $T_\mathrm{w}$.
        }
\end{figure}
In \S \ref{s:scatter} we define \tmin/, the best-fitting boxcar-average equivalent time-scale for ensembles of points in SFHs grouped in moving windows of fiducial width $T_\mathrm{w}$ = 500 Myr. 
In Figure \ref{f:window_test} we show that the choice of $T_\mathrm{w}$, for sufficiently large $T_\mathrm{w}$, does not affect the inferred average value of \tmin/.
For \halpha/ (bottom panel), \tmin/ is largely insensitive to the window size so long as it is larger than $T_\mathrm{w}= 50$ Myr  (save for the first 1 Gyr, which is affected by edge effects near $t=0$).  
On the other hand, FUV (top panel) shows stronger variations for $T_\mathrm{w}<250$ Myr, but above this value \tmin/ is well converged.




\bibliographystyle{mnras}
\bibliography{references} 


\bsp	
\label{lastpage}
\end{document}